\pdfoutput=1
\documentclass[useAMS,usenatbib,usegraphicx]{mn2e}

\usepackage{amsmath,graphicx,amssymb,natbib}
\usepackage{times}

\newcommand\hvezda{{$\sigma$~Ori~E}}
\newcommand{\zav}[1]{\left(#1\right)}

\newcommand\intvidpo{\!\!\int\limits_{\begin{array}{c}\text{\scriptsize
visible}\\[-2mm]\text{\scriptsize surface}\end{array}}\!\!}

\newlength\staretab

\makeatletter
\def\sgn{\mathop{\operator@font sgn}\nolimits}
\makeatother

\DeclareMathAlphabet{\mathsc}{OT1}{cmr}{m}{sc}
\def\testbx{bx}%
\DeclareRobustCommand{\ion}[2]{%
\relax\ifmmode
\ifx\testbx\f@series
{\mathbf{#1\,\mathsc{#2}}}\else
{\mathrm{#1\,\mathsc{#2}}}\fi
\else\textup{#1\,{\mdseries\textsc{#2}}}%
\fi}

\title[Revisiting the RRM model for $\sigma$ Ori E II.]{Revisiting the Rigidly Rotating Magnetosphere model for $\sigma$ Ori E - II. Magnetic Doppler imaging, arbitrary field RRM, and light variability\thanks{Based on observations obtained using the Narval spectropolarimeter at the Observatoire du Pic du Midi (France), which is operated by the Institut National des Sciences de l'Univers (INSU) and observations obtained at the Canada-France-Hawaii Telescope (CFHT) which is operated by the National Research Council of Canada, the Institut National des Sciences de l'Univers of the Centre National de la Recherche Scientifique of France, and the University of Hawaii}}
\author[M.E. Oksala et al.]{M. E. Oksala$^{1,2,3}$\thanks{E-mail:meo@udel.edu}
O. Kochukhov$^{4}$, J. Krti\v{c}ka$^{5}$, R. H. D. Townsend$^{6}$, G. A. Wade$^{7}$,   
\newauthor  M. Prv\'ak$^{5}$, Z. Mikul\'a\v{s}ek$^{5}$, J. Silvester$^{7,8}$, S. P. Owocki$^{1}$ \\
$^{1}$Bartol Research Institute, Department of Physics and Astronomy, University of Delaware, Newark, DE 19716, USA\\
$^{2}$Astronomical Institute, Academy of Sciences of the Czech Republic, Fricova 298, 251 65 Ond\v{r}ejov, Czech Republic \\
$^{3}$LESIA, Observatoire de Paris, CNRS UMR 8109, UPMC, Universit\'{e} Paris Diderot, 5 place Jules Janssen, 92190, Meudon, France \\
$^{4}$Department of Physics and Astronomy, Uppsala University, Box 516, Uppsala 75120, Sweden \\
$^{5}$Institute of Theoretical Physics and Astrophysics, Masaryk University, 611 37 Brno, Czech Republic \\\
$^{6}$Department of Astronomy, University of Wisconsin-Madison, 2535 Sterling Hall, 475 N Charter Street, Madison, WI 53706, USA \\
$^{7}$Department of Physics, Royal Military College of Canada, P.O. Box 17000, Station Forces, Kingston, Ontario K7K 7B4, Canada\\
$^{8}$Department of Physics, Engineering Physics \& Astronomy, Queen'€™s University, Kingston, Ontario K7L 3N6, Canada \\
}
\begin{document}

\date{\today}

\pagerange{\pageref{firstpage}--\pageref{lastpage}} \pubyear{2014}

\maketitle

\label{firstpage}

\begin{abstract}

The initial success of the Rigidly Rotating Magnetosphere (RRM) model application to the B2Vp star $\sigma$~Ori~E by \citet{town05} triggered a renewed
era of observational monitoring of this archetypal object.   {We utilize high-resolution spectropolarimetry and the magnetic Doppler imaging (MDI)
technique to simultaneously determine the magnetic configuration, which is predominately dipolar, with a polar strength B$_{\rm{d}} = 7.3-7.8$ kG and a smaller non-axisymmetric quadrupolar contribution, as well as the surface distribution of abundance of He, Fe, C, and Si.  }
We describe a revised RRM model that now accepts an arbitrary surface magnetic field configuration, with the field topology from the MDI models used as input.  The resulting synthetic H$\alpha$ emission and broadband photometric observations generally agree with observations, however, several features are poorly fit.
To explore the possibility of a photospheric contribution to the observed photometric variability, the MDI abundance maps 
were used to compute a synthetic photospheric light curve to determine the effect of the surface inhomogeneities.
Including the computed photospheric brightness modulation fails to improve the agreement between the observed and computed 
photometry. We conclude that the discrepancies cannot be explained as an effect of inhomogeneous surface abundance.  
Analysis of the UV light variability shows good agreement between observed variability and computed light curves, 
supporting the accuracy of the photospheric light variation calculation.
We thus conclude that significant additional physics is necessary for the RRM model to acceptably reproduce observations of not only $\sigma$~Ori~E, 
but also other similar stars with significant stellar wind-magnetic field interactions.

\end{abstract}


\section{Introduction}

\begin{figure}
\centering
\includegraphics[width=90mm]{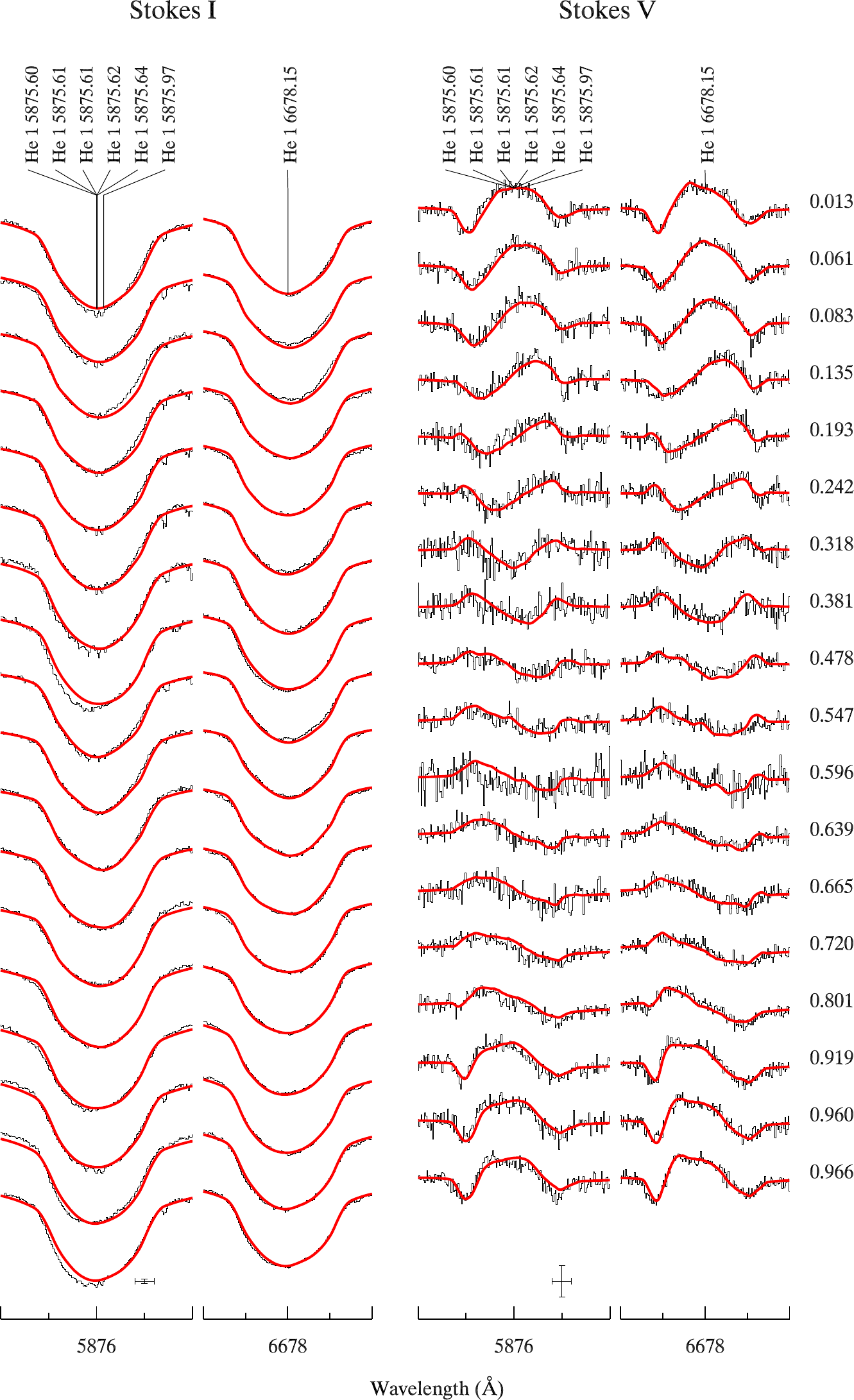}
\caption{Comparison between observed (black) and computed (red) Stokes $I$ and $V$ profiles of the He\,{\sc i} 5876\AA\ and 6678\AA\ lines.
Rotational phases are indicated to the right of the Stokes $V$ profiles.  Spectra corresponding to consecutive phases are shifted vertically. 
The bars at the lower left corner of each panel indicate the horizontal (1 \AA) and vertical (5 per cent) scales of the profile plots.}
\label{MDImagprof}
\end{figure}

A small percentage of main sequence A- and B-type (Ap/Bp) stars are characterized by peculiar surface abundances of chemical elements and strong (0.1-30 kG), 
large-scale magnetic fields.  These magnetic Ap/Bp stars exhibit periodic variability of spectral lines, photometric 
brightness, and longitudinal magnetic field strength that is explained in the context of the Oblique Rotator Model (ORM) described by 
\citet{stibbs}.
The peculiar element abundances present themselves in the form of inhomogeneous, asymmetric ``spots'' on the 
surface of the star, causing spectral line shape and strength variability, and in some cases photometric brightness variations, 
all modulated on the stellar rotation period.  
These inhomogeneities on the surface of magnetic chemically peculiar stars are thought to be 
formed by direct or indirect interactions between chemical diffusion and the stabilizing influence of magnetic field lines \citep[see e.g.,][]{michaud}.

The technique of Doppler imaging (DI) was originally used by \citet{gonchar} to map the abundance of a chemical 
element on the surface of a star assuming rotational variability of line profiles with time.
This simple, yet elegant method of studying stellar spots was eventually extended to simultaneously map both chemical abundances as well as a star's surface magnetic field.
This method, magnetic Doppler imaging (MDI), was previously employed to study Ap/Bp stars, and
successfully investigates both the surface inhomogeneities of various elements and the magnetic field topology.  
The MDI code {\sevensize INVERS}10 was developed by \citet{PK02} to accurately model stellar magnetic fields 
and surface abundance features from an inversion of high resolution spectropolarimetry. 
The code was originally developed to operate with data in all four Stokes parameters.  However, it is also able to handle a set of data comprised of
only Stokes $I$ and $V$ spectra.

Using abundance maps derived from DI, \citet{Krt07} showed that both the photometric and spectroscopic 
variability of another He-strong star, HD~37776, can be explained by the presence of inhomogeneous abundances across the stellar surface.  Similar work 
corroborates this claim for the case of the rapidly rotating Ap star CU Vir \citep{Krt12}.   
\citet{Krt13} determined that the photospheric spots of HD~64740 are not sufficient to cause detectable photometric variability.  
Further, these authors identified UV line variability due to these spots, without any evidence of circumstellar features.

\begin{figure*}
\centering
\includegraphics[width=150mm]{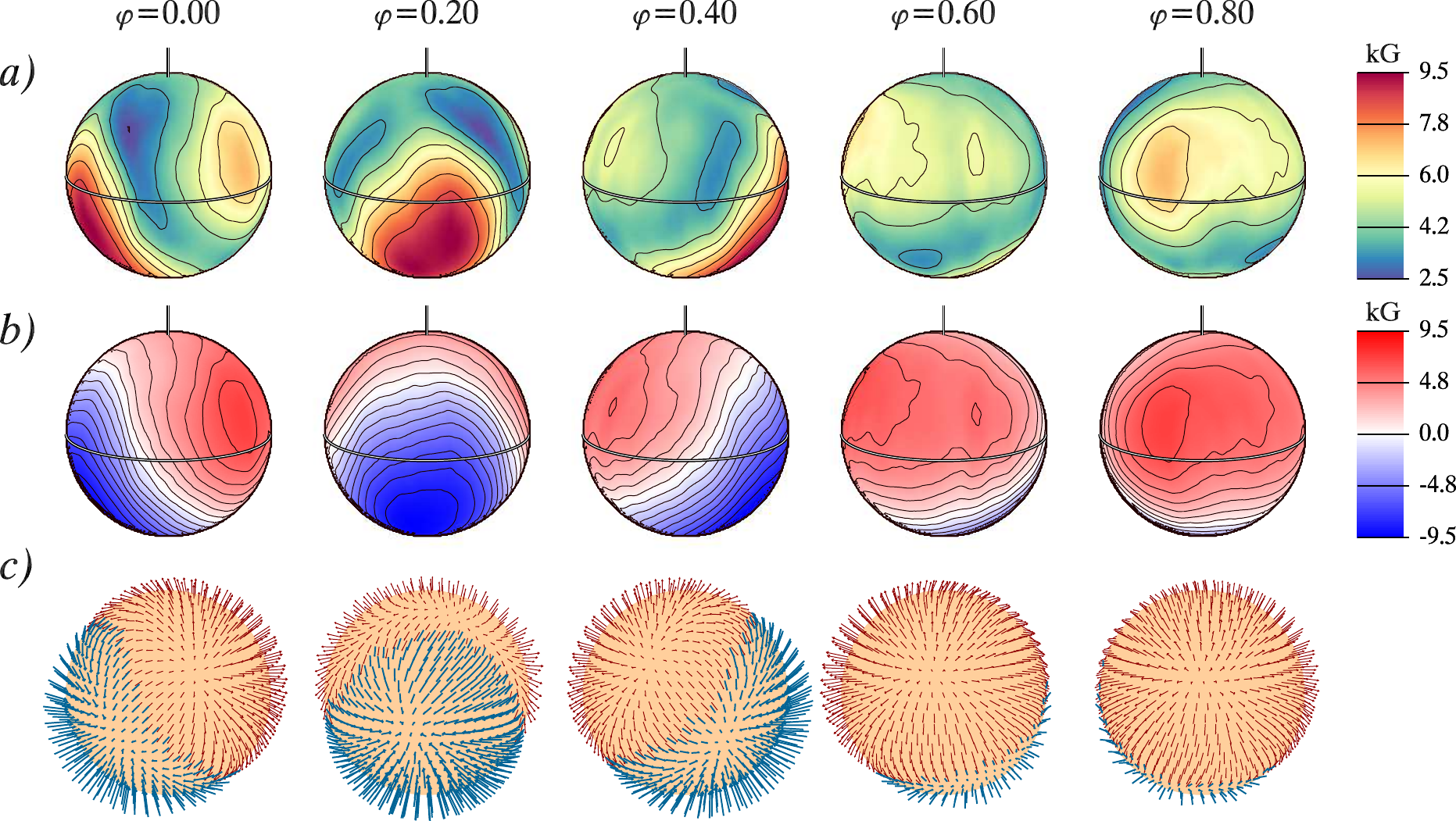}
\caption{Surface magnetic field distribution derived for $\sigma$ Ori E with magnetic DI.  
The star is displayed at five equidistant rotational phases as indicated at the top of the figure.  
The rotational axis is indicated by the short vertical bar, while the thick line shows the stellar rotational equator. 
The maps correspond to an inclination angle $i = 75$\degr and a $v_\text{rot} \sin i =  140$ km s$^{-1}$. 
\textbf{a):} Maps of the magnetic field modulus.   The contour lines are plotted with a step of 1.0 kG. 
\textbf{b):} Maps of the radial magnetic field strength.  The contour lines are also plotted with a step of 1.0 kG. 
\textbf{c):} Orientation of magnetic field vectors.  In the vector maps, light (red) arrows point outward from the 
stellar surface.  Dark (blue) arrows correspond to vectors pointing inwards.  The arrow length is proportional to the field strength.}
\label{MDImagmap}
\end{figure*}

$\sigma$ Orionis E (HD~37479), the prototypical magnetic He-strong Bp star, has long been known to exhibit variable line profiles, 
indicating an inhomogeneous surface distribution of various elements.  {The discovery of its magnetic field by \citet{LB78}
suggested that this star was a higher mass extension of the Ap/Bp phenomena, and could be physically represented by the ORM.}
A study by \citet{R00} modeled historical and new
spectra to investigate line profile variations to determine their relationship to local differences of element abundances.  
The authors found regions of He overabundance and metal (C~{\sc ii} and Si~{\sc iii}) underabundance slightly offset from each other, 
suggested to be located at the magnetic poles.  

\citet{Walborn74} had discovered broad, variable H$\alpha$ emission features, which, when considered in the context of the ORM, indicated 
that the emitting material was magnetospheric in nature \citep{LB78}. 
Various authors worked to develop realistic magnetospheric models for this and other
similar stars, with the most recent being the development of the Rigidly Rotating Magnetosphere (RRM) model by \citet{TO05}.  
The model analytically describes a rapidly rotating star surrounded by plasma trapped in its magnetosphere.  Although the RRM model
was relatively successful in reproducing the magnetospheric spectral and photometric variability of $\sigma$ Ori E \citep{town05}, major inconsistencies remained.

To begin addressing these issues and the revision of the RRM model,
\citet[][hereafter referred to as Paper I]{Oksala} presented and analyzed high resolution intensity and circular polarization spectra 
$\sigma$ Ori E.  These spectra identified various spectral lines that exhibited significant variability and presented an updated 
magnetic field characterization, suggesting a more complex, higher order field structure. This result was in strong contradiction with the offset 
dipole configuration adopted for the specific case of $\sigma$ Ori E by \citet{town05}.

A remaining issue with the RRM model, which we aim to address here, are persistent discrepancies between the observed optical photometric light curve and the
model light curve (see for example Figure 8, this paper).  As noted by \citet{town05}, a significant observed feature of increased brightness is unmatched at rotational phase 0.6 (see their Fig.~1). 
As these authors invoked the de-centered dipole as a means to mimic the unequal depths of the light minima, 
a change of the magnetic configuration would cause changes to the synthetic light curve, and further differences will likely arise in the 
model-observation comparison.  The most precise photometric data to date were obtained over a period of three weeks with the \textit{MOST} micro-satellite, 
and were presented and analyzed by \citet{town13}.  The emission
and absorption features previously seen in the light curve of $\sigma$ Ori E are clearly present, and support the notion that these features are real, and also stable. 
The properties of the two minima strongly differ, with one eclipse deeper than the other, while the more shallow feature is present during a longer portion of the rotation period.  The excess brightness at  phase 0.6 is quite evident, and the high precision reveals a gradual decline of this excess before returning to the ``continuum'' level.

This subsequent work continues our re-evaluation of the RRM model, utilizing cutting edge techniques to analyze the magnetic and abundance
structure of $\sigma$ Ori E via MDI, revise the RRM model, and, through a thorough light curve analysis, address any remaining discrepancies  
between model and observation.  Section 2 explains the MDI procedure and results.  Section 3 presents the revised RRM model.  We discuss in Section 4 the light curve analysis of the optical photometry.  An analysis of UV variability is explored in Section 5.  Section 6 presents a summary of this paper.

\section{Magnetic field and elemental abundance mapping}

\subsection{Magnetic Doppler imaging (MDI)}

For input into MDI, we obtained a total of 18 high-resolution (R=65000) broadband (370-1040~nm) Stokes $I$ and $V$ spectra of $\sigma$ Ori E.  
Sixteen spectra were obtained in November 2007 with the Narval spectropolarimeter on the 2.2m Bernard Lyot 
telescope (TBL) at the Pic du Midi Observatory in France.  The remaining two spectra were obtained in February 
2009 with the spectropolarimeter ESPaDOnS on the 3.6-m Canada-France-Hawaii Telescope (CFHT), as part of the 
Magnetism in Massive Stars (MiMeS) Large Program \citep{Wade}.  The observations and their reduction are described 
in more detail in Paper I.

\begin{figure}
\centering
\includegraphics[width=80mm]{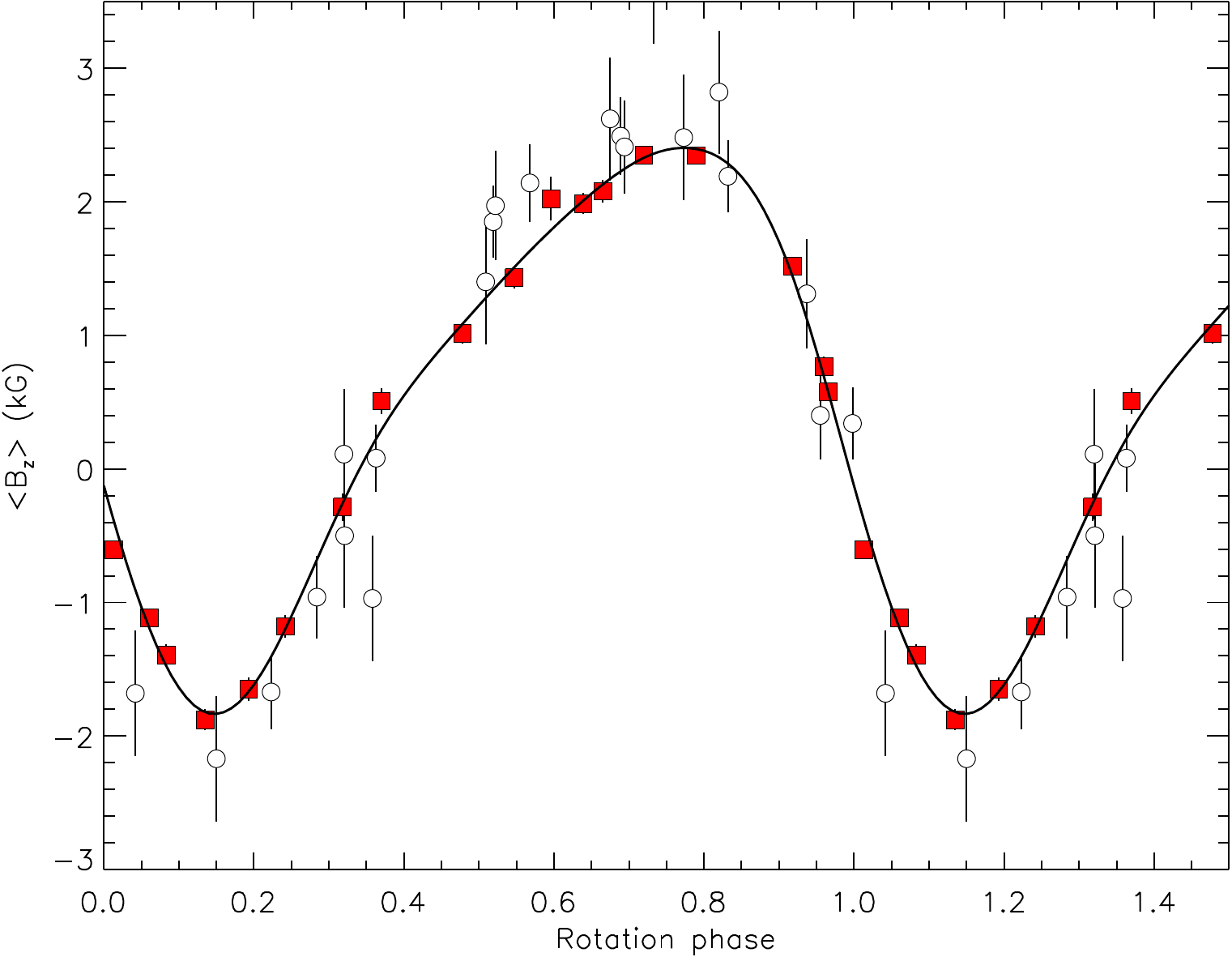} 
\caption{Longitudinal magnetic field measurements (red filled squares) for $\sigma$ Ori E (Paper I), as well as data reported in Landstreet \& Borra (1978)  and Bohlender et al. (1987) (open circles). Each data point is plotted with corresponding 1$\sigma$ error bars. The solid curve is the longitudinal magnetic field curve computed using the MDI-derived magnetic field structure. The data are phased using the ephemeris of Townsend et al. (2010).}
\label{Bzcurve}
\end{figure}

\begin{figure*}
\centering
\includegraphics[width=150mm]{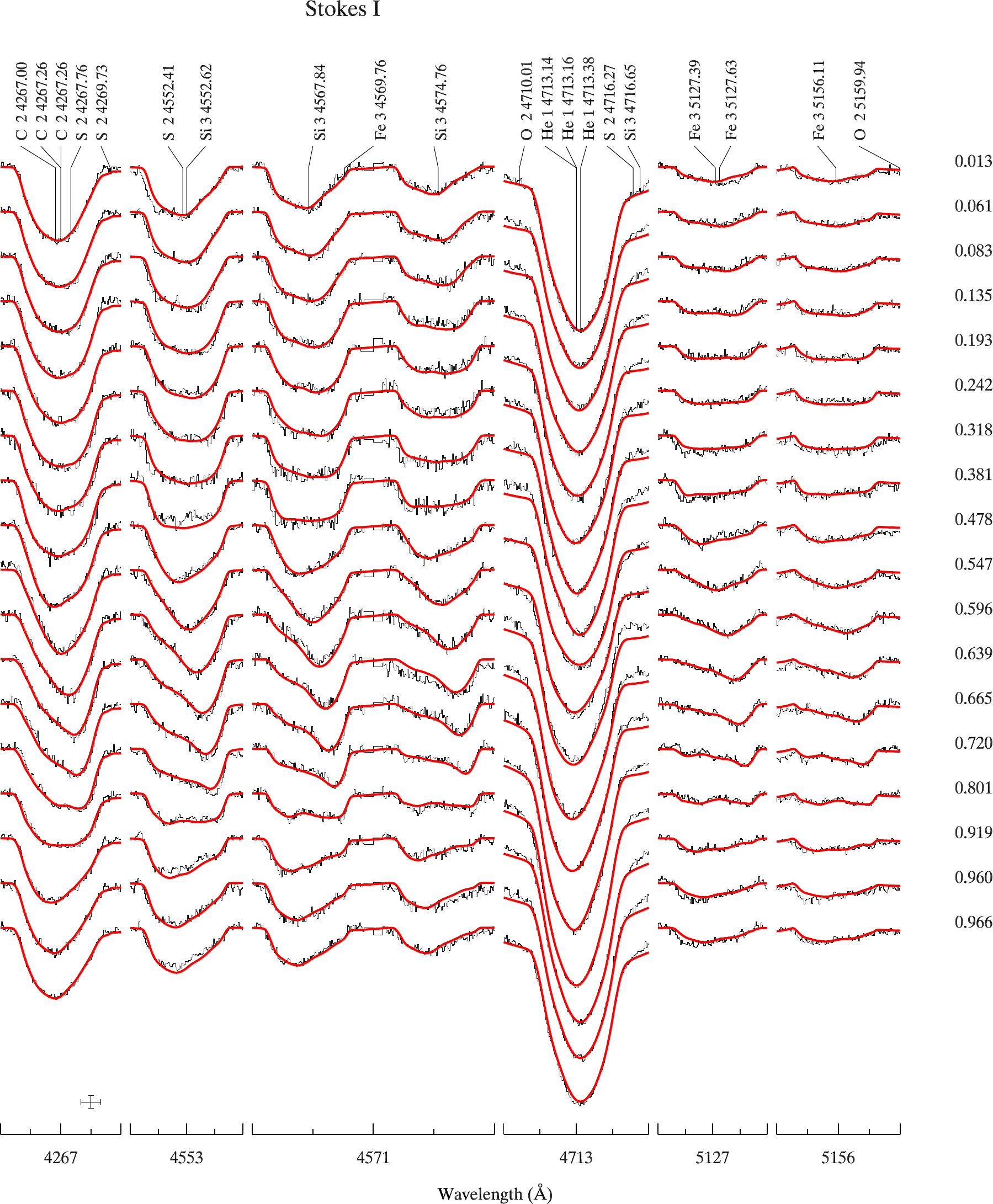} 
\caption{Comparison between observed (black) and computed (red) Stokes $I$ profiles of lines used for chemical abundance mapping..
Rotational phases are indicated to the right of the profiles.  Spectra corresponding to consecutive phases are shifted vertically. 
The bars at the lower left corner of each panel indicate the horizontal (1 \AA) and vertical (5 per cent) scales of the profile plots.}
\label{MDIabnprf}
\end{figure*}

\begin{figure*}
\centering
\includegraphics[width=150mm]{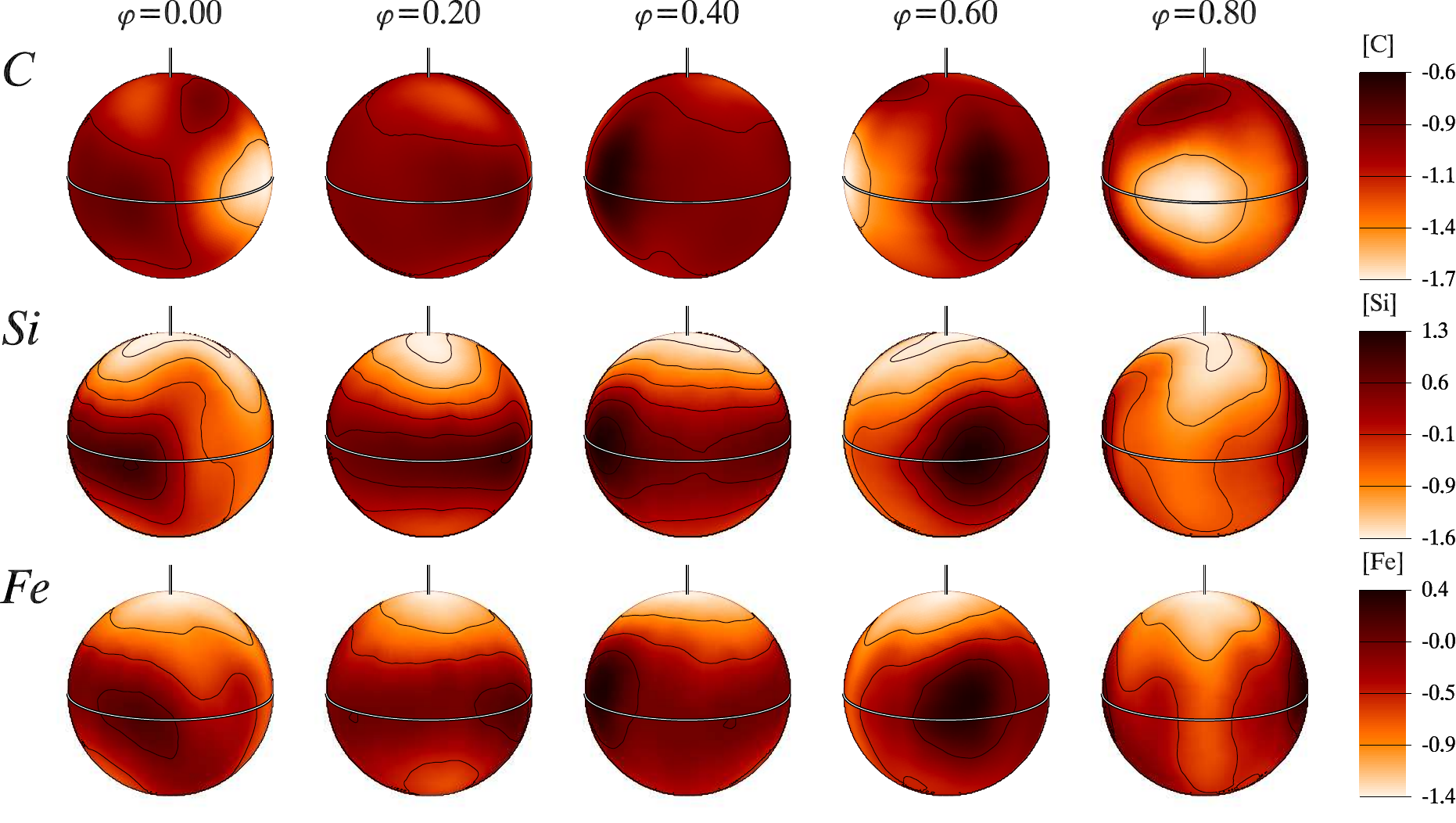} 
\caption{The chemical abundance distributions of C, Fe, and Si derived from Stokes $I$ line profiles. 
The star is shown at 5 equidistant rotational phases viewed at the inclination
angle, $i = 75\degr$ and $v \sin i = 140$ km s$^{-1}$. The scale gives abundance as $\epsilon_{Elem}$ corresponding to
$\log(\rm{N}_{Elem} /\rm{N}_{tot})$ relative to solar values. The rotation axis is vertical.  The contour step size is 0.5 dex.
}
\label{MDImetalabnmap}
\end{figure*}

\begin{figure*}
\centering
\includegraphics[width=150mm]{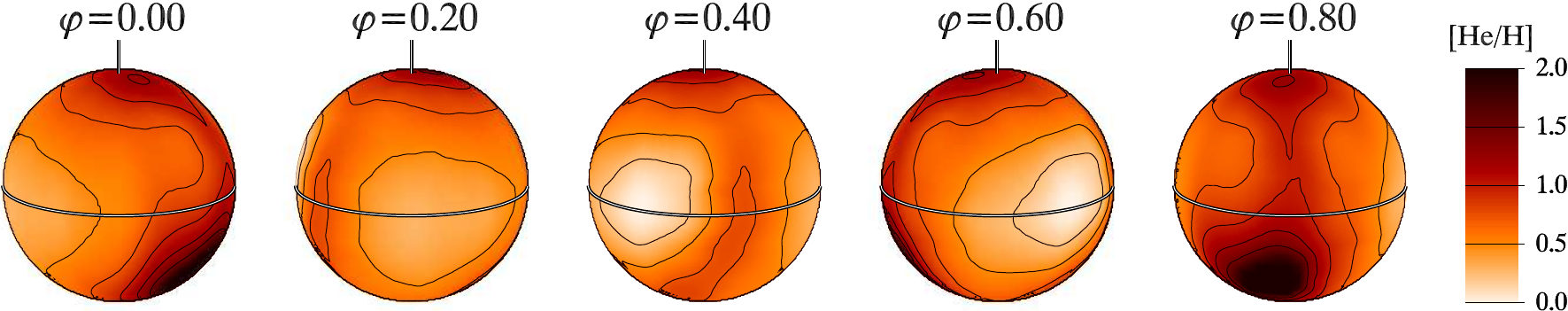}  
\caption{The chemical abundance distribution of He derived from Stokes $I$ line profiles of
the 4713 \AA\ He\,{\sc i} line. The star is shown at 5 equidistant rotational phases viewed at the inclination
angle, $i = 75\degr$ and $v \sin i = 140$ km s$^{-1}$. The scale gives abundance as $\log(\rm{He} /\rm{H})$ relative to solar values. 
The rotation axis is vertical.  The contour step size is 0.5 dex.
}
\label{MDIHeabnmap}
\end{figure*}

In this work, we utilize the {\sevensize INVERS}10 MDI code \citep{PK02} and its recent derivative {\sevensize INVERS}13 \citep{Koch12,Koch13} to reconstruct the 
surface abundance distribution and magnetic field topology of $\sigma$ Ori E using 
these high-resolution Stokes $I$ and $V$ spectra. Both codes use least-squares
minimization together with a regularization procedure to fit synthetic spectra to spectropolarimetric observations. The
model spectra are adjusted until the solution converges to the simplest surface distribution that properly describes the observational data.  

Because only Stokes $I$ and $V$ spectra were obtained, the final magnetic field solution is not unique,
and the choice of regularization becomes important. {As described by \citet{PK02}, without Q and U spectra, we must use an additional a priori constraint, multipolar regularization, 
which constrains the results according to agreement with a best fit multipolar model magnetic field.  
Additionally, Stokes $IV$ observations only provide information about the global magnetic field geometry, and is thus unable to reveal small-scale magnetic field structures, such as those seen in some full Stokes $IQUV$ inversions of Ap stars \citep[e.g.,][]{Koch04,KW10, SKW14}}
We also include the longitudinal field curve as an additional constraint in magnetic inversion following the procedure described by \citet{Koch02}.
{The resulting field configuration is computed as part of the multipolar regularization procedure.  Thus, the multipolar parameters reported here represent an approximation of the  2-D MDI magnetic map, which is derived without using the spherical harmonic parameterization. }

\begin{figure*}
\centering
\includegraphics{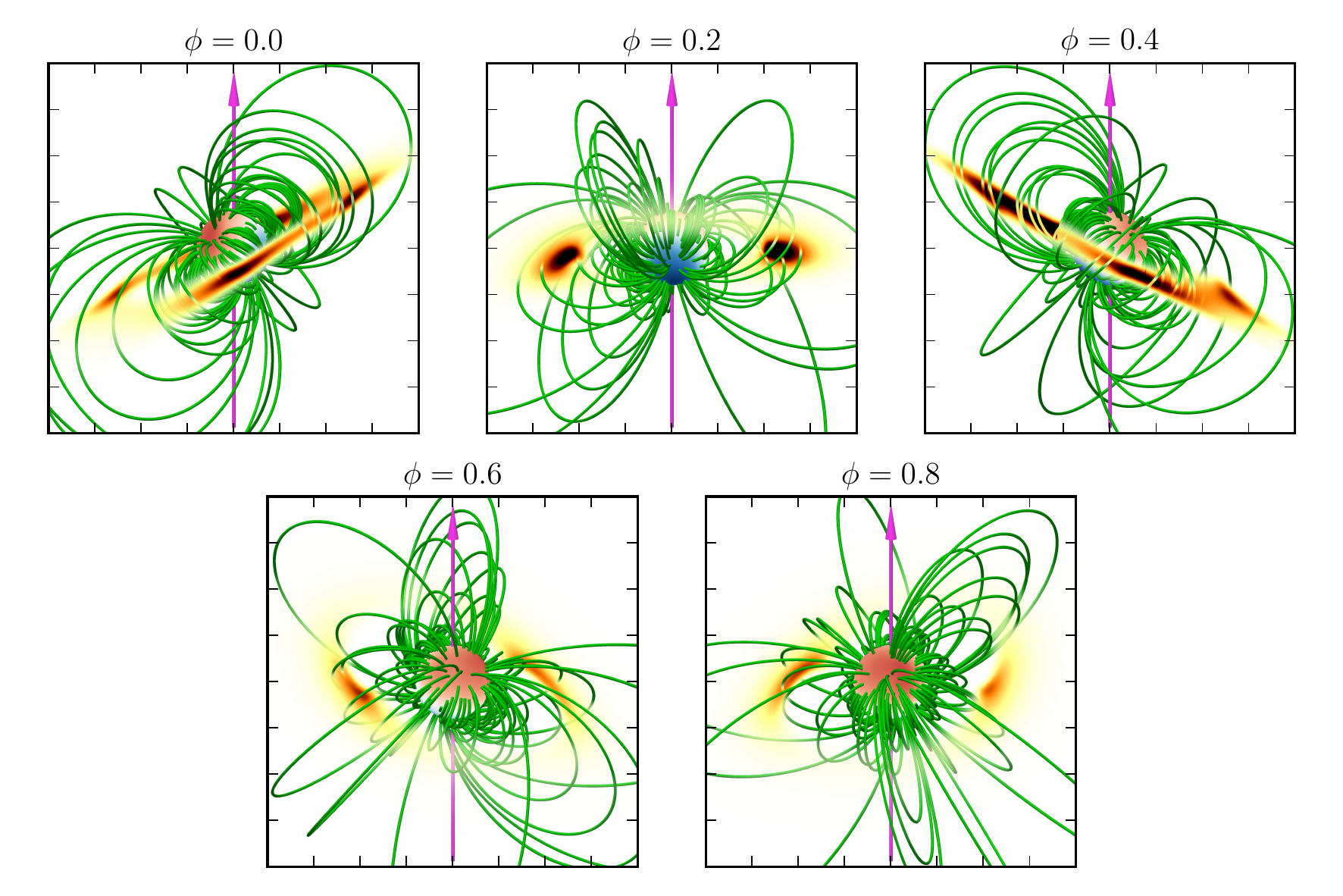}  \\
\caption{Visualizations of the plasma distribution predicted by the
  ARRM model, at the same 5 rotational phases adopted in previous
  figures. The star is colored according to the radial magnetic field
  strength, as in the middle panels of Fig.~\ref{MDImagmap}; selected
  field lines are plotted in green, and the magenta arrow shows the
  star's rotation axis.  The surrounding distribution of magnetospheric material is shown via an orange-colored volume rendering, with the more opaque parts corresponding to higher column densities and vice-versa.  The tick marks on the surrounding boxes are spaced by $1\,R_*$.}
\label{fig:arrm-vis}
\end{figure*}

\subsection{Magnetic field topology}\label{Bfield}

To determine the magnetic field topology of $\sigma$ Ori E, the line profiles of the 
He\,{\sc i} lines at 5876 \AA\ and 6678 \AA\ were chosen for the inversion process.  These lines both have strong magnetic 
signatures in their Stokes $V$ spectrum, while other possible lines exhibit much weaker polarization features, and are therefore less suitable for use in magnetic inversion.
We use as input the parameters (${{T}_\mathrm{eff}}$, ${\log g}$, and $v_\text{rot} \sin i$) determined in Paper I, 
which are listed in Table~\ref{hvezda}.  
The rotational period, $P_{\rm{rot}}$, was derived by \citet{town10}.
{While these parameters are rather well established for $\sigma$ Ori E, there is a large uncertainty in the determination of the inclination angle, $i$.  This is a result of 
uncertainty in the size and distance of the star, and as such the radius.  A range of radius values have been reported via various methods, and so we compute MDI maps at 4 fixed 
inclination angles to understand the effect of this parameter.  Corresponding to a range in radii of 3.3-4.0 R$_\star$, models were computed for $i$ = 55, 65, 75, and 85\degr.  This range is adopted based on the discussion of the stellar radius in the Appendix of \citet{town13}, accounting for approximate uncertainties in the angular diameter \citep{GH82} and the cluster distance \citep{Sherry}.}

While the {\sevensize INVERS}10 code relies on LTE model atmospheres, this assumption is not always realistic.  For hot stars, the LTE assumption 
begins to break down starting at T$_{\rm{eff}} > 15000$ K \citep{MihalAthay}, beginning with visibly different line profile 
strengths and/or shapes. These effects become particularly noticeable above T$_{\rm{eff}} \sim 22000$ K, resulting in growing 
discrepancies with LTE model spectra in H\,{\sc i} lines and He\,{\sc i} lines, particularly those in the red-ward part of
the optical spectrum \citep[i.e., 5876 and 6678 \AA;][]{Przy}.  Less obvious effects are seen in metal lines and blue-ward 
He~{\sc i} lines; the study by \citet{Przy} finds little difference between LTE and NLTE models for these lines.  
Thus, to properly model the He\,{\sc i} lines we have chosen, these changes 
need to be incorporated into the synthetic spectrum.  The implementation of full NLTE radiative transfer into the INVERS 
suite of codes is not a trivial task, and we proceed to recover the line profile shapes only
for lines to be included in the magnetic field determination.
The effect of NLTE departure is more severe for red He\,{\sc i} lines, however, the magnetic signature is also 
stronger in these features, making them useful for determining the magnetic
configuration.  As a solution, we have used the NLTE model atmosphere code
TLUSTY \citep{bstar2006} to determine the departure coefficients for the upper and lower atomic levels of the He transitions in question.
The code was set to only consider particle conservation and statistical equilibrium to calculate both the population levels and their associated 
departure coefficients.  Departure coefficients were computed for various values of helium abundance with a LTE ATLAS9 \citep{kurucz} model atmosphere with
T$_{\rm{eff}} = 23000$~K and ${\log g}={4.0}$ as the initial input for the TLUSTY computation. 
Then, during inversion, {\sevensize INVERS}10 interpolated departure coefficients for the local value of He abundance and modified the line absorption coefficient and the source function accordingly.

With the longitudinal magnetic field curve derived in Paper I as an additional constraint in addition to the multipolar 
regularization \citep{PK02}, we obtained a good fit to both the Stokes $I$ and $V$ line profiles of the two He\,{\sc i} lines, shown in Figure\,\ref{MDImagprof}.  {These figures correspond to an inclination of $i$ = 75\degr.
The best fit magnetic field configuration over the considered range of inclination angles ($i = 55-85$\degr)
suggests a magnetic field that can be approximated as a dipole with polar strength $B_{\rm{d}} = 7.3-7.8$ kG, with obliquity $\beta_{\rm{d}} = 47-59$\degr\,
and phase angle $\gamma_{\rm{d}} = 97-98$\degr, and a smaller non-axisymmetric quadrupole component with strength $B_{\rm{q}} = 3-5$ kG.  Figure~\ref{MDImagmap} shows the surface magnetic
field distribution maps derived for $\sigma$ Ori E corresponding to this topology (again, for $i = 75$\degr). 
The dipole is centered, but the quadrupole axis is misaligned with the dipolar axis, leading to a geometry where the positive and negative poles are clearly not separated by 180\degr. }  The asymmetry is further
evident in the longitudinal field curve shown in Figure~\ref{Bzcurve}, where the observed longitudinal field measurements from all epochs are 
plotted along with the synthetic curve computed from the MDI magnetic maps.  The data are very well matched, and corroborate
the conclusions of Paper I, particularly their Figure 7, which shows a comparatively poor fit to the observed data using the original RRM de-centered dipole field configuration.

\subsection{Metal abundance}
\label{metalabun}

{With a magnetic field topology corresponding to an inclination $i = 75$\degr 
set as a fixed parameter, {\sevensize INVERS}10 was then used to compute the distribution 
of various chemical elements on the surface of the star.    
This inclination was chosen as it was used for the original $\sigma$ Ori E RRM analysis by \citet{town05}, 
and we expect far less uncertainty in the determination of the surface fields with varying inclination angle.}
The set of high resolution time-series spectra of $\sigma$ Ori E contain lines suitable for modeling 
the abundances of Fe, Si, C, and He.  Other chemical elements are present in the spectrum, but their 
line strength was too weak for accurate determination of any surface structure.
He lines required a slightly more rigorous treatment, and will be discussed in the next section. 
  
The Stokes $I$ spectral lines were fit, with the final results shown in Figure~\ref{MDIabnprf}.  There is good agreement 
between the observations and the model fits for all of these lines.  Figure~\ref{MDImetalabnmap} displays 
spherical maps of the abundance distributions of Fe, Si, and C.   
The maps of Si abundance were determined using three closely located Si\,{\sc iii} lines at 
4552.62 \AA, 4567.84 \AA, and 4574.76 \AA. The surface distribution of C was computed from the C\,{\sc ii} 4267 \AA\,triplet.
Fe maps were produced using the Fe\,{\sc iii} doublet lines at 5127.39 and 5127.63 \AA\,and the Fe\,{\sc iii} singlet line at 5156.11 \AA.
The surface maps shown in Figure~\ref{MDImetalabnmap} indicate that Fe, Si, and C 
all show a similar pattern over the stellar surface.  The minimum abundances are found at 
rotational phase 0.8, in both a spot at the equator and at the visible pole.  
An equatorial spot at $\sim$ phase 0.6 has the maximum value on the surface.

\begin{figure}
\centering
\includegraphics{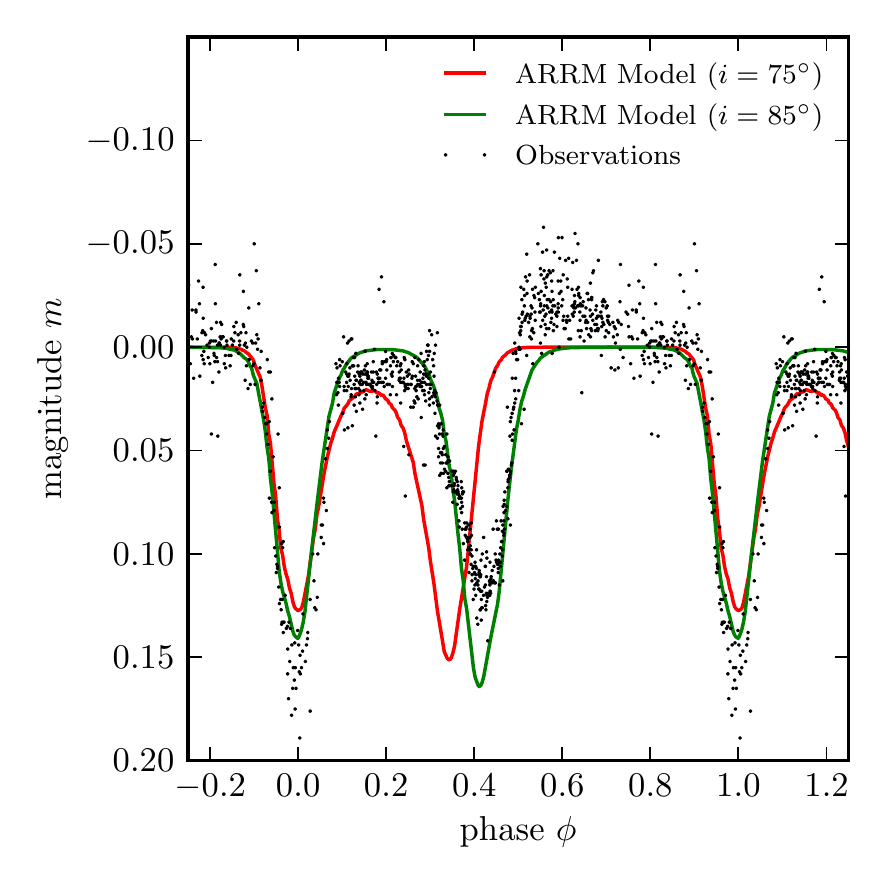}  \\
\caption{Comparison between the observed \citep[black dots;][]{H77} and modeled photometric light curves for $i = 75\degr$ (red curve) and $i = 85\degr$ (green curve), for the Str\"omgren $u$-band.}
\label{fig:arrm-light}
\end{figure}

\subsection{He abundance}
\label{Heabun}

As discussed in Section~\ref{Bfield}, the two helium lines used for magnetic mapping are more sensitive 
to departure from NLTE.  Further, these lines are very strong, and as such they have a weaker ability to diagnose 
changes in abundance.  We find that He\,{\sc i} 5876 \AA\,and 6678 \AA\,also contain contamination from the circumstellar
 material, which was confirmed in Paper I by evaluating the variability of the equivalent widths (EWs) of these lines.
We therefore choose to determine the photospheric helium abundance using the weaker He\,{\sc i} 4713 \AA\,line,
which does not show any distortion due to either NLTE effects or the magnetosphere.  Further, instead of using the standard {\sevensize INVERS}10
MDI code to compute synthetic He line profiles, we chose to use a different imaging code,
{\sevensize INVERS}13 \citep[for more details see][]{Koch12,Koch13}. This MDI code is capable of computing local Stokes parameter profiles 
according to local atmospheric conditions, modified by the presence of chemical or temperature spots. In the case of $\sigma$ Ori E, which shows a large He overabundance
and a non-uniform surface distribution of this element, we computed a grid of LTE model atmospheres with LLmodels code \citep{llmodels} for a range of
He/H ratios. This grid was then used in modeling of the He\,{\sc i} 4713 \AA\ line, allowing us to derive a more accurate He surface map than when using a single mean atmosphere.
As with the metals, the He abundance distribution was computed 
with the fixed magnetic field geometry determined from the red He lines.

Figure~\ref{MDIHeabnmap} shows a He abundance map indicating a large spot of overabundance at phase 0.8, located in the lower hemisphere.
This corresponds with the EW variation in the He\,{\sc i} 6678 and 5876 \AA\ lines derived in Paper I.
The minimum abundance is located at phase 0.6, with a normal, solar level of He.  The spot of enhanced He does not appear
to be correlated with the location of the magnetic poles.  Also of note is the anti-correlation between the strength of metals
and He, previously established in Paper I from EW variations.

\begin{figure*}
\centering
\includegraphics{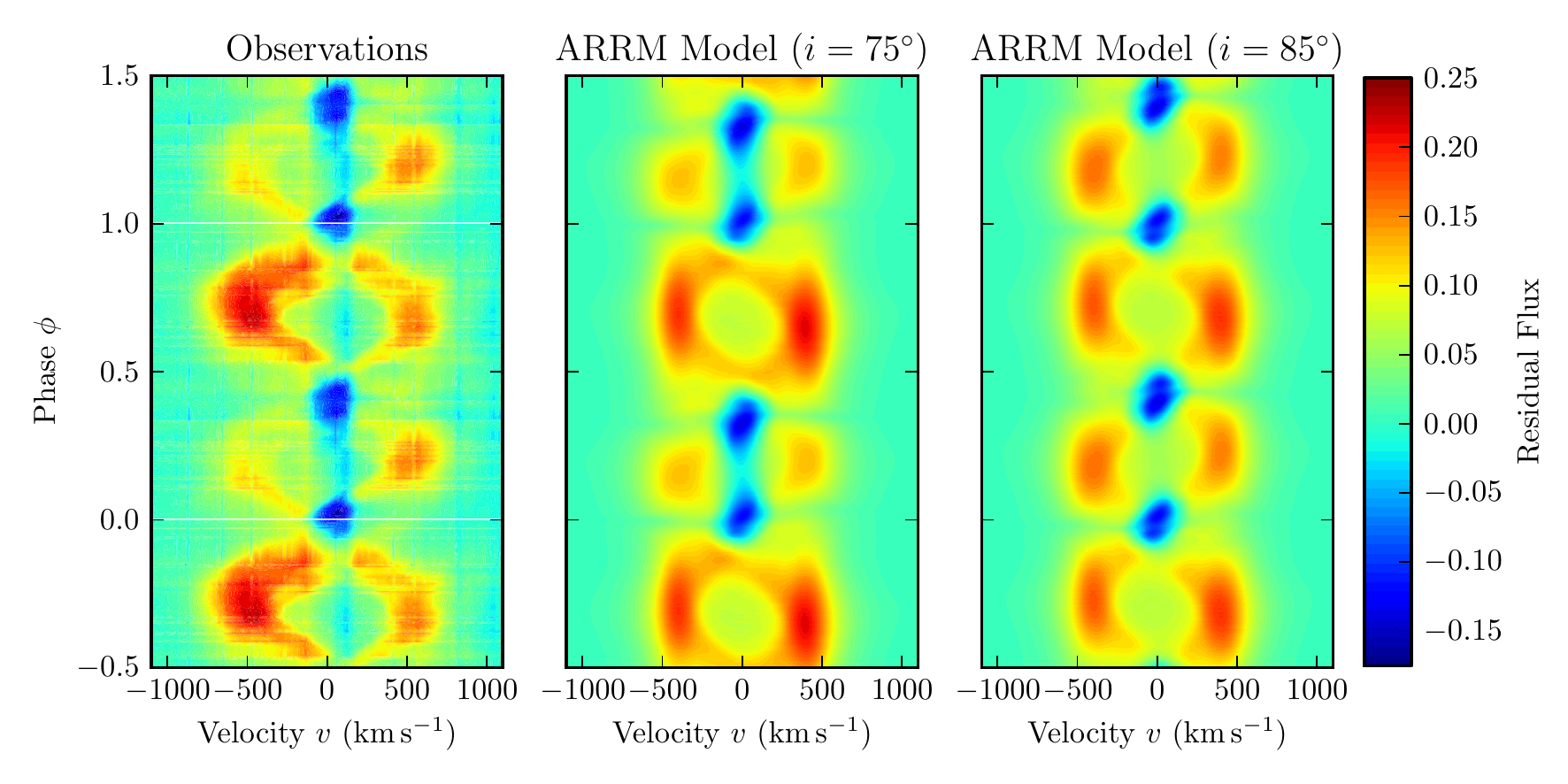}  \\
\caption{Comparison between observed (left) and modeled (middle and right) H$\alpha$ dynamic spectra of $\sigma$~Ori~E, 
in each case plotting the residual flux as a function of velocity and rotation phase. The middle plot is determined for an inclination angle, $i = 75\degr$, while the right plot is the same, but for $i = 85\degr$.  
The observational data and the synthetic reference spectrum used to calculate the residuals are described in Section 4 of paper I.}
\label{fig:arrm-halpha}
\end{figure*}

\section{The revised Rigidly Rotating Magnetosphere model for $\sigma$ Ori  E}

The initial success of applying the Rigidly Rotating Magnetosphere
(RRM) model to $\sigma$~Ori~E, by \citet{town05}, triggered a renewed
era of observational monitoring of this archetypal object, across a
range of different wavelengths.  In Paper I, we presented the
first-ever high-resolution spectropolarimetry of $\sigma$~Ori~E, which
reveals that the star's magnetic topology is more complex than assumed
in the original RRM analysis. Here, we explore the consequences of
this revised topology, considering the analysis of Section~\ref{Bfield}, 
for the RRM-predicted distribution of circumstellar material.

\subsection{The Offset-Dipole RRM Model}

The RRM analysis by \citet{town05} assumed a dipole magnetic topology
tilted at an angle $\beta = 55^{\circ}$ with respect to the rotation
axis and then offset by $0.3\,R_*$ in a direction essentially
perpendicular to both rotational and magnetic axes. The offset results
in an mass/density asymmetry in the predicted magnetospheric mass
distribution (with respect to the plane containing both axes), which
nicely explains the observed inequalities in the depths of the
light-curve minima and the heights of the H$\alpha$ emission peaks.

\citet{town05} reported that their offset-dipole topology produces a
reasonable fit (reduced $\chi^{2} = 1.79$) to the 
longitudinal magnetic field ($B_{\ell}$) measurements available at that time,
only marginally worse than a centered dipole. However, a review of the
field synthesis code used by these authors has revealed a couple of
bugs, which tend to suppress the effects of the offset on the
$B_{\ell}$ curve. With these bugs fixed, the predicted $B_{\ell}$ is
clearly inconsistent with the observations: the transition from
magnetic maximum to minimum is gradual, while the transition from minimum of maximum is quite steep -- precisely
the opposite of the observed behavior (see Fig.~3).

This mismatch could be partly mitigated by offsetting the dipole in
the \emph{opposite} direction (at the expense of worsening the light
curve and H$\alpha$ fits). However, recognizing that an offset dipole
is at best a crude approximation to the star's true magnetic topology,
and that we now have access to {a more physically realistic topology} (Sec.\ref{Bfield}),
a better approach is to discard the offset dipole approach and
instead extend the RRM model to handle arbitrary non-dipolar fields.

\subsection{A New Arbitrary-Field RRM Model}

The Arbitrary-Field RRM (ARRM) model is an extension of the original
RRM formalism to allow consideration of arbitrary circumstellar field
topologies. The theoretical basis remains the same: the distribution
of density $\rho$ along a given closed magnetic flux tube is
calculated by solving the equation of hydrostatic equilibrium subject
to gravitational and centrifugal forces. This yields
\begin{equation} \label{eqn:rho-s}
\rho(s) = 
\begin{cases}
\rho_{0} \exp \left[ -\mu \frac{\Phi(s) - \Phi_{0}}{kT} \right] & s_{a} < s < s_{b} \\
0 & \text{otherwise}
\end{cases},
\end{equation}
where $s$ is the coordinate measuring arc distance along the flux
tube; $s_{a}$ and $s_{b}$ are the coordinates of the maxima in the
effective potential $\Phi(s)$ that are situated closest to the tube
footpoints, and the other symbols have the same meaning as described by
\citet{TO05}. In this expression, which is equivalent to equation~(25)
of \citet{TO05}, $\rho_{0} \equiv \rho(s_{0})$ is the density at an
arbitrary reference coordinate $s=s_{0}$, and $\Phi_{0}\equiv
\Phi(s_{0})$ is the corresponding effective potential.

To determine $\rho_{0}$, we integrate the density
distribution~(\ref{eqn:rho-s}) along the full length of the flux tube,
giving the total mass $m$ in the tube as
\begin{equation} \label{eqn:m}
m = \int_{s_{\rm S}}^{s_{\rm N}} \rho(s)\, {\rm d}A(s)\,{\rm d}s.
\end{equation}
Here, ${\rm d}A(s)$ is the nominal cross-sectional area of the tube,
and $s_{\rm N}$ and $s_{\rm S}$ are the arc coordinates of the
northern ($B_{r} > 0$) and southern ($B_{r} < 0$) tube footpoints,
respectively. Conservation of magnetic flux requires that
\begin{equation}
|\mathbf{B}(s)| \, {\rm d}A(s) = |\mathbf{B}(s_{\rm N})| \, {\rm d}A(s_{\rm
  N}) = |\mathbf{B}(s_{\rm S})| \, {\rm d}A(s_{\rm S}).  
\end{equation}
Likewise, conservation of mass requires that
\begin{equation}
 m = \dot{m} t,
\end{equation}
where $\dot{m}$ is the rate at which mass is being added to the tube,
and $t$ is the time elapsed since the magnetosphere last began to
refill\footnote{Recently, \citet{town13} have presented evidence that
  the process(es) responsible for limiting the total mass in a
  magnetosphere may be more complex than the episodic centrifugal
  breakouts envisaged in the appendix of \citet{TO05}; however, in the
  present analysis we assume that the latter picture remains
  correct.}. Generalizing equation~(31) of \citet{TO05}, the filling
rate $\dot{m}$ can be related to the global CAK mass-loss rate
$\dot{M}$ via
\begin{equation}
\dot{m} = \frac{\dot{M}}{4\pi R_*} \left( \frac{B_{r}(s_{\rm N})}{|\mathbf{B}(s_{\rm N})|}\,{\rm d}A(s_{\rm N}) + \frac{B_{r}(s_{\rm S})}{|\mathbf{B}(s_{\rm S})|}\,{\rm d}A(s_{\rm S}) \right)
\end{equation}
With the arbitrarily chosen timescale $t$ setting the overall mass contained in
the magnetosphere, and for given stellar parameters and magnetic
topology, these equations suffice to calculate the density everywhere
in the magnetosphere.

\begin{figure}
\centering \resizebox{\hsize}{!}{\includegraphics{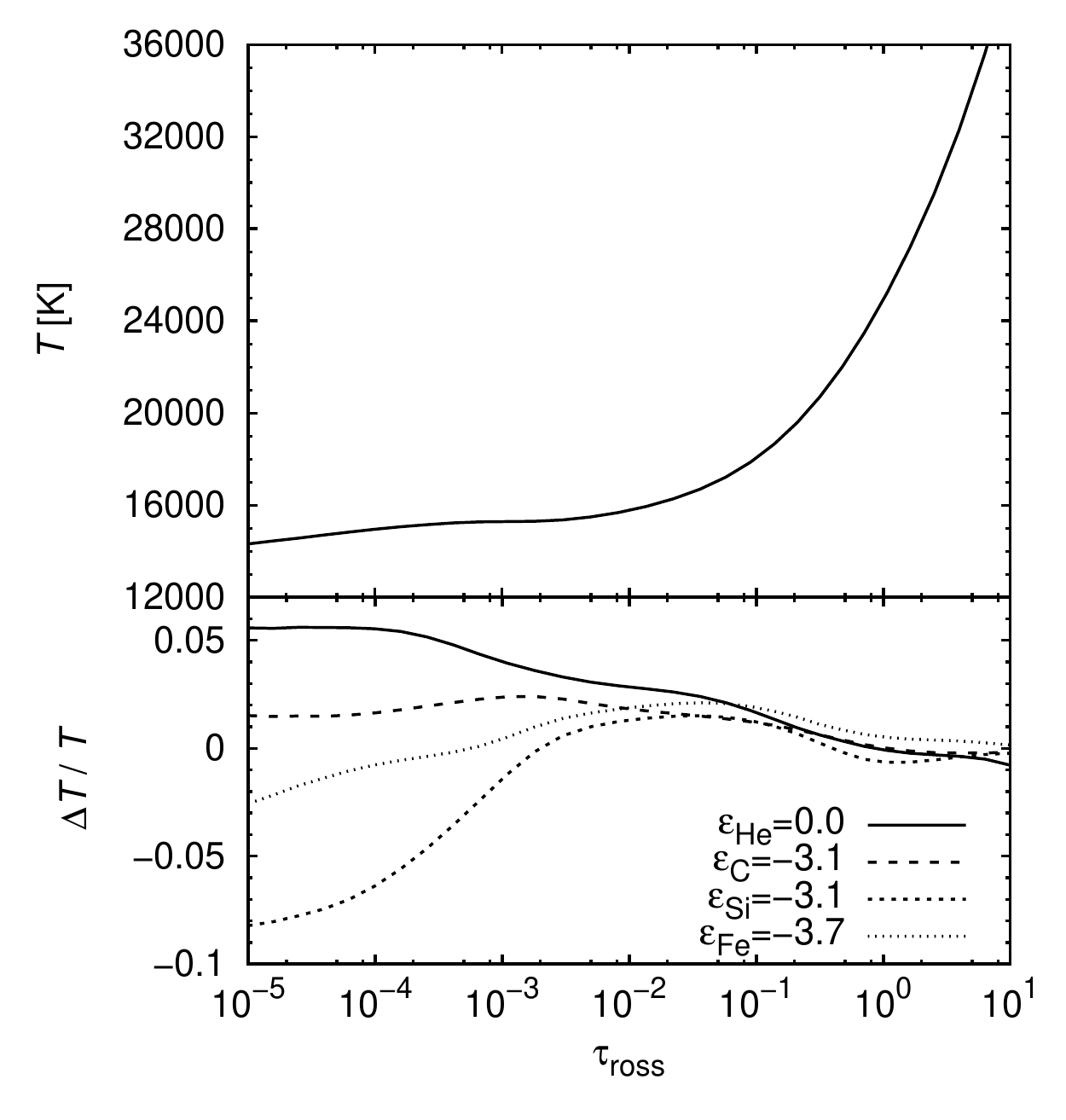}}
\caption{{\em Upper plot:} The dependence of temperature on the Rosseland
optical depth $\tau_\text{ross}$ in
a reference model atmosphere with a slightly overabundant chemical composition ($\varepsilon_\text{He}=-1.0$,
$\varepsilon_\text{C}=-4.1$, $\varepsilon_\text{Si}=-4.1$, and
$\varepsilon_\text{Fe}=-4.7$). {\em Lower plot}: The relative change of the
temperature in the model atmospheres with modified abundance.}
\label{tep}
\end{figure}

\subsection{Applying the ARRM Model to $\sigma$~Ori~E} \label{resultsARRM}

\begin{figure*}
\centering \resizebox{0.75\hsize}{!}{\includegraphics{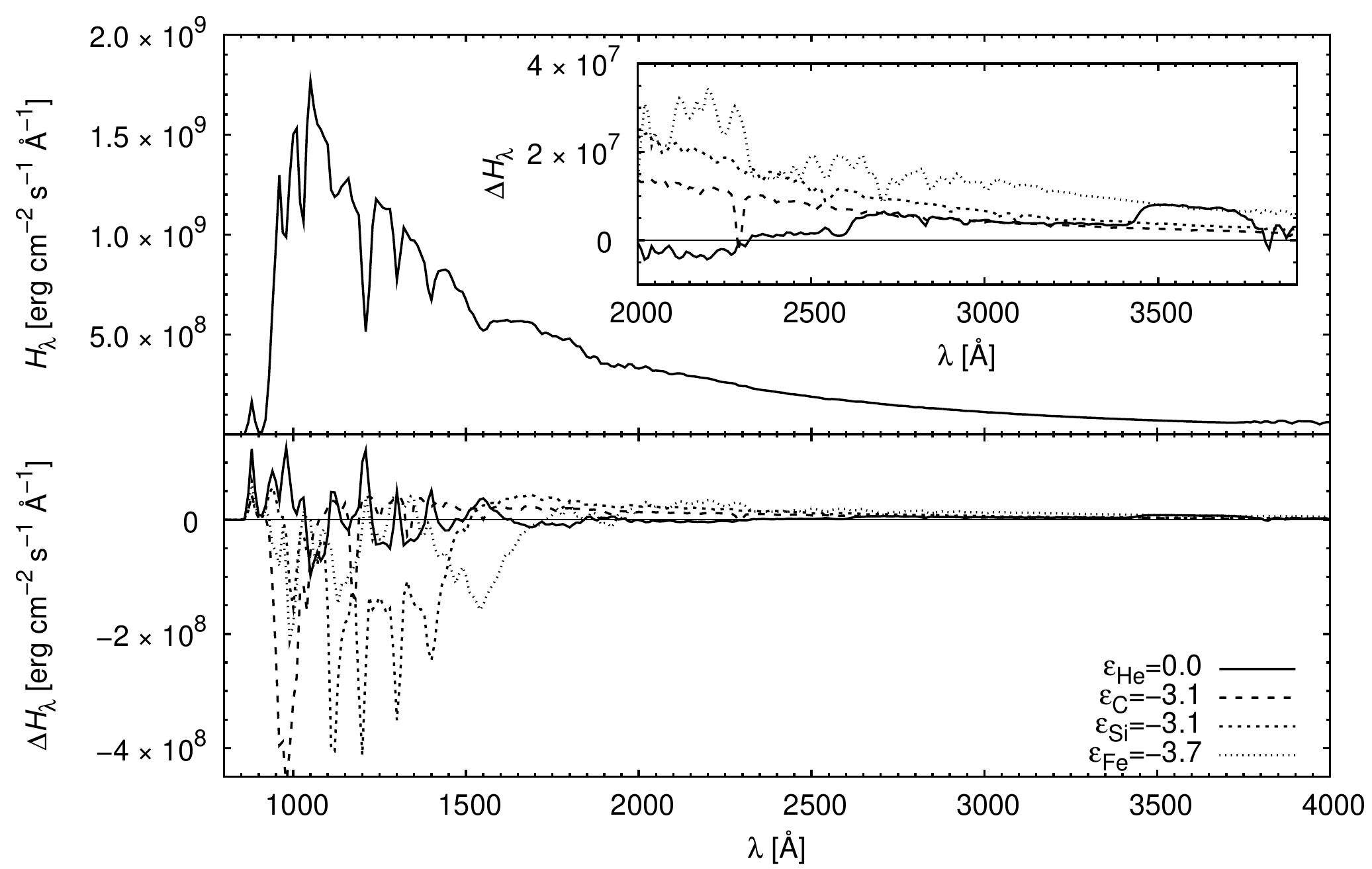}}
\caption{{\em Upper plot:} The emergent flux from a reference model
atmosphere with slightly overabundant chemical composition 
($\varepsilon_\text{He}=-1.0$, $\varepsilon_\text{C}=-4.1$,
$\varepsilon_\text{Si}=-4.1$, and $\varepsilon_\text{Fe}=-4.7$). {\em Lower
plot}: The emergent flux from the model atmospheres with modified abundance of
individual elements minus the flux from the reference model. A magnified section of the lower plot
 showing the redistribution of flux to longer wavelengths is over-plotted in the small upper panel. All fluxes were
smoothed by a Gaussian filter with a dispersion of $10\,$\AA\ to show the
changes in continuum, a major contribution to SED variability.
}
\label{prvtoky}
\end{figure*}

Applying the ARRM model to $\sigma$~Ori~E requires a
prescription for the magnetic field vector $\mathbf{B}$ throughout the
magnetosphere. To this end, we extrapolate from the surface radial
field distribution using the potential-based approach described by
\citet{jard02}.  {This method is employed with the assumption that the field remains strongly dominant 
over the wind to a distance within which the majority of the material is located.  
As can be seen directly in Fig. 9, the material responsible for the H$\alpha$ emission is located in a radius range of $\sim$2-4 R$_*$.
Since the density is a strong function of radius, we expect the amount of material in the magnetosphere to greatly decrease with distance from the star.  This agrees with the bulk of the material being located close to the Keplerian co-rotation radius, which is consistent with the results of MHD simulations by \citet{uddoula}.  Moreover, at this distance, the magnetic tension dominates the centrifugal force by a factor of $\sim10^3$, assuring minimal distortion from a pure dipole form.  The determination of where the balance between these forces 
becomes more equal, and how it determines the size and shape of the disc are issues that, along with further details about the implementation of this extrapolation, will be given in a forthcoming paper (Townsend et al., in prep.).}

Figure~\ref{fig:arrm-vis} presents visualizations of the magnetic
topology and the associated density distribution predicted by the ARRM
model for $\sigma$~Ori~E, at the same five rotation phases illustrated
in Figs.~2 and~4, for an inclination angle $i = 75\degr$. 
The material in the magnetosphere is distributed in
an oblique, warped disk-like structure, qualitatively the same as
shown Fig.~4 of \citet{town05} for a simple oblique dipole. This is
not unexpected --- the rapid falloff of higher-order components means
that the field in the ARRM model, at the distances $r \gtrsim
2\,R_{*}$ where the material is situated, is close to dipolar.
{The model simulations keep free the parameter which designates the total amount 
of material in the magnetosphere, so that the observations can be best matched.  The model
used here indicates a mass density within the magnetosphere of $5\times10^{-12}$ g cm$^{-3}$, 
or a number density of $5\times10^{12}$ cm$^{-3}$. }

The light curve and H$\alpha$ dynamic spectrum predicted by the ARRM
model are compared against the corresponding observational data in
Figs.~\ref{fig:arrm-light} and~\ref{fig:arrm-halpha}.   The photometric data are 
 Str\"omgren $u$ band taken by \citet{H77}.  The observed H$\alpha$ spectra
 are part of the data presented here and in Paper I.  
The approach used to calculate these observables is similar to that described by
\citet{town05}, except that we now include an NLTE photospheric line profile
(as described in Sec. 4.1 of Paper I) in the spectral synthesis rather
than a flat continuum.  {The comparison has been shown for both $i = 75$ and $85\degr$; we do not show 
 similar plots for $i = 55$ or $65\degr$, as they make the comparison much worse, and
 as such can be ruled out as possible inclination angles.   
 Clearly, the model predictions computed with $i = 85\degr$ more closely match the timing
 of the observed variability, however, a clear difference between observations and
model can be seen in Fig.~\ref{fig:arrm-light}: the secondary minimum
at $\phi \sim 0.4$ is deeper than the primary minimum at $\phi \sim 0.0$,
the opposite of what is seen in the observations. }This reversal is also
apparent in the strength of the H$\alpha$ emission peaks, with the
blue peak being stronger than the red one at $\phi \sim 0.75$ in the
observations, but vice-versa for the model.  {We also note that the models do not extend the H$\alpha$
emission out far enough in velocity, particularly on the blue side, something that will be explored in the next iteration of this model.
Furthermore, the improvements added to develop the ARRM model are not sufficient to explain the enhanced emission at phase 0.6.
}




\section{Photospheric contribution to the optical brightness variation}

The previous section (Sec.\ref{resultsARRM}) demonstrates that the current version of the RRM model, the ARRM model,
is not effective at explaining the totality of the features seen in the optical variability of $\sigma$ Ori E.  However, the model 
is strictly accounting for the circumstellar contribution to these observables, without consideration of any other possible contribution to the variability.  In the following sections, we
explore the contribution from the stellar photosphere, accounting for the effects of an inhomogeneous abundance distributions of several elements.

\subsection{Simulation of the SED variability}

\begin{table}
\caption{Adopted physical parameters of \hvezda.  Table references: $^1$\citet{Oksala}, $^2$\citet{town10}.
}
\label{hvezda}
\begin{center}
\begin{tabular}{lc}
\hline
Effective temperature ${{T}_\mathrm{eff}}$$^1$ & ${23\,000}\pm3\,000$\,K \\
Surface gravity ${\log g}$ (cgs)$^1$ & ${4.0}\pm0.5$ \\
Projected rotational velocity $v_\text{rot} \sin i$$^1$ & $140\pm10\,\text{km}\,\text{s}^{-1}$\\
Rotational period P$_{\rm{rot}}$$^2$ & $1\fd1908229 + 1\fd44\times10^{-9}E$ \\
Helium abundance&$-1.1<\varepsilon_\text{He}<0.6$ \\
Carbon abundance& $-5.0<\varepsilon_\text{C}<-4.0$ \\
Silicon abundance& $-5.9<\varepsilon_\text{Si}<-3.2$ \\
Iron abundance&$-5.7<\varepsilon_\text{Fe}<-4.0$  \\
\hline
\end{tabular}
\end{center}
\end{table}

\subsubsection{Model atmospheres and synthetic spectra}

The simulation of the ultraviolet and visual SED variability primarily utilizes the
NLTE model atmosphere code TLUSTY \citep{tlusty,hublaj,hublad,lahub}. 
The atomic data \citep[adopted from][]{bstar2006} are appropriate for
B-type stars. These data were originally calculated within the Opacity and Iron
Projects \citep{topt,zel0}. The calculations assume fixed values of the
effective temperature and surface gravity for $\sigma$ Ori E
(according to Table~\ref{hvezda}), and
adopt a generic value of the microturbulent velocity
$v_\text{turb}=2\,\text{km}\,\text{s}^{-1}$. The abundances of helium, carbon,
silicon, and iron were taken from the four-parameter abundance grid given in
Table~\ref{esit} covering the range of abundances found on the \hvezda\ surface
as inferred from the MDI mapping discussed in Sections\,\ref{metalabun} and \ref{Heabun}.
Abundances relative to hydrogen were used, i.e., $\varepsilon_\text{el}=\log\zav{N_\text{el}/N_\text{H}}$.
All other elements were set to their solar \citep{asgres} values.

Synthetic spectra were calculated using the SYNSPEC code 
for the same parameters (effective temperature, surface
gravity, and abundances) and transitions as the model atmosphere calculations. Additionally, we
included lines of all elements with atomic number $Z\leq30$ not included
in the model atmosphere calculation. Angle-dependent intensities,
$I(\lambda,\theta,\varepsilon_\text{He},\varepsilon_\text{C},\varepsilon_\text{Si},\varepsilon_\text{Fe})$,
were computed for $20$ equidistantly spaced values of $\cos\theta$, where $\theta$ is the
angle between the normal to the surface and the line of sight.

The generation of the complete grid of the model atmospheres and angle-dependent
intensities (Table~\ref{esit}) would require calculation of 700 model
atmospheres and synthetic spectra. Because not all abundance combinations correspond
to physical conditions determined by the abundance maps, we calculated only those
models that were realistic. This helped us to reduce the number of calculated
models by about two thirds.

\begin{table}
\caption{Individual abundances $\varepsilon_\text{He}$, $\varepsilon_\text{C}$,
$\varepsilon_\text{Si}$, and $\varepsilon_\text{Fe}$ of the model grid}
\label{esit}
\begin{center}
\begin{tabular}{lrrrrrrr}
\hline
He& $-1.0$& $-0.5$& $0.0$& $0.5$& $1.0$\\
C& $-5.1$& $-4.6$& $-4.1$& $-3.6$ \\
Si& $-6.1$& $-5.6$& $-5.1$& $-4.6$& $-4.1$& $-3.6$& $-3.1$ \\
Fe & $-5.7$& $-5.2$& $-4.7$& $-4.2$& $-3.7$ \\
\hline
\end{tabular}
\end{center}
\end{table}

\begin{figure}
\centering
\resizebox{\hsize}{!}{\includegraphics{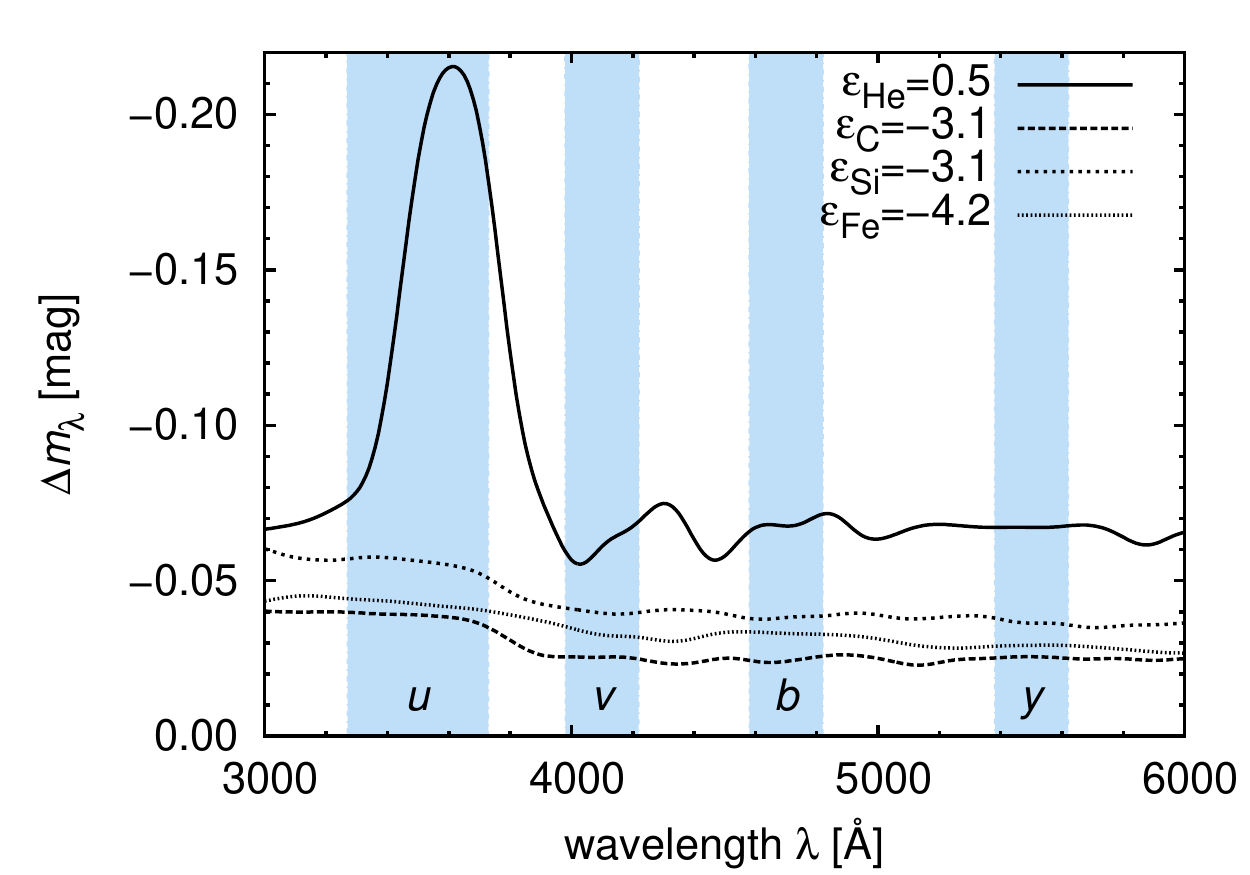}}
\caption{The magnitude difference $\Delta m_\lambda$ between the emergent fluxes
calculated with enhanced abundance of individual elements and the reference flux
$H_\lambda^\text{ref}$ (see Eq.~\ref{tokmagroz}).
The fluxes were smoothed by a Gaussian filter with a dispersion
of $100\,$\AA.}
\label{magtoky}
\end{figure}

\subsubsection{Calculation of phase dependent SED}
\label{vypocet}

The radiative flux transmitted through a photometric band $c$ at a distance $D$ from a star with
radius $R_*$ is \citep{mihalas}
\begin{equation}
\label{vyptok}
f_c=\zav{\frac{R_*}{D}}^2\intvidpo I_c(\theta,\Omega)\cos\theta\,\text{d}\Omega.
\end{equation}
The specific band intensity $I_c(\theta,\Omega)$ is obtained by interpolating between intensities
$I_c(\theta,\varepsilon_\text{He},\varepsilon_\text{C},\varepsilon_\text{Si},\varepsilon_\text{Fe})$ 
for abundances found at the surface point with spherical coordinates $\Omega$.
These intensities are calculated from the grid of synthetic spectra (see
Table~\ref{esit}) according to
\begin{equation}
\label{barint}
I_c(\theta,\varepsilon_\text{He},\varepsilon_\text{C},\varepsilon_\text{Si},\varepsilon_\text{Fe})=
\int_0^{\infty}\Phi_c(\lambda) \,
I(\lambda,\theta,\varepsilon_\text{He},\varepsilon_\text{C},\varepsilon_\text{Si},\varepsilon_\text{Fe})\, \text{d}\lambda.
\end{equation}
The response function $\Phi_c(\lambda)$ for individual bands is
derived by either fitting the tabulated response functions or simply
assuming a Gaussian function.

The magnitude difference in a given band is defined as
\begin{equation}
\label{velik}
\Delta m_{c}=-2.5\,\log\,\zav{\frac{{f_c}}{f_c^\mathrm{ref}}},
\end{equation}
where $f_c$ is calculated from Eq.~\ref{vyptok} and
${f_c^\mathrm{ref}}$ is the reference flux obtained under the
condition that the mean magnitude difference over the rotational period is zero.

\subsection{Influence of abundance on emergent flux}
\label{kaptoky}

\begin{figure}
\centering
\resizebox{\hsize}{!}{\includegraphics{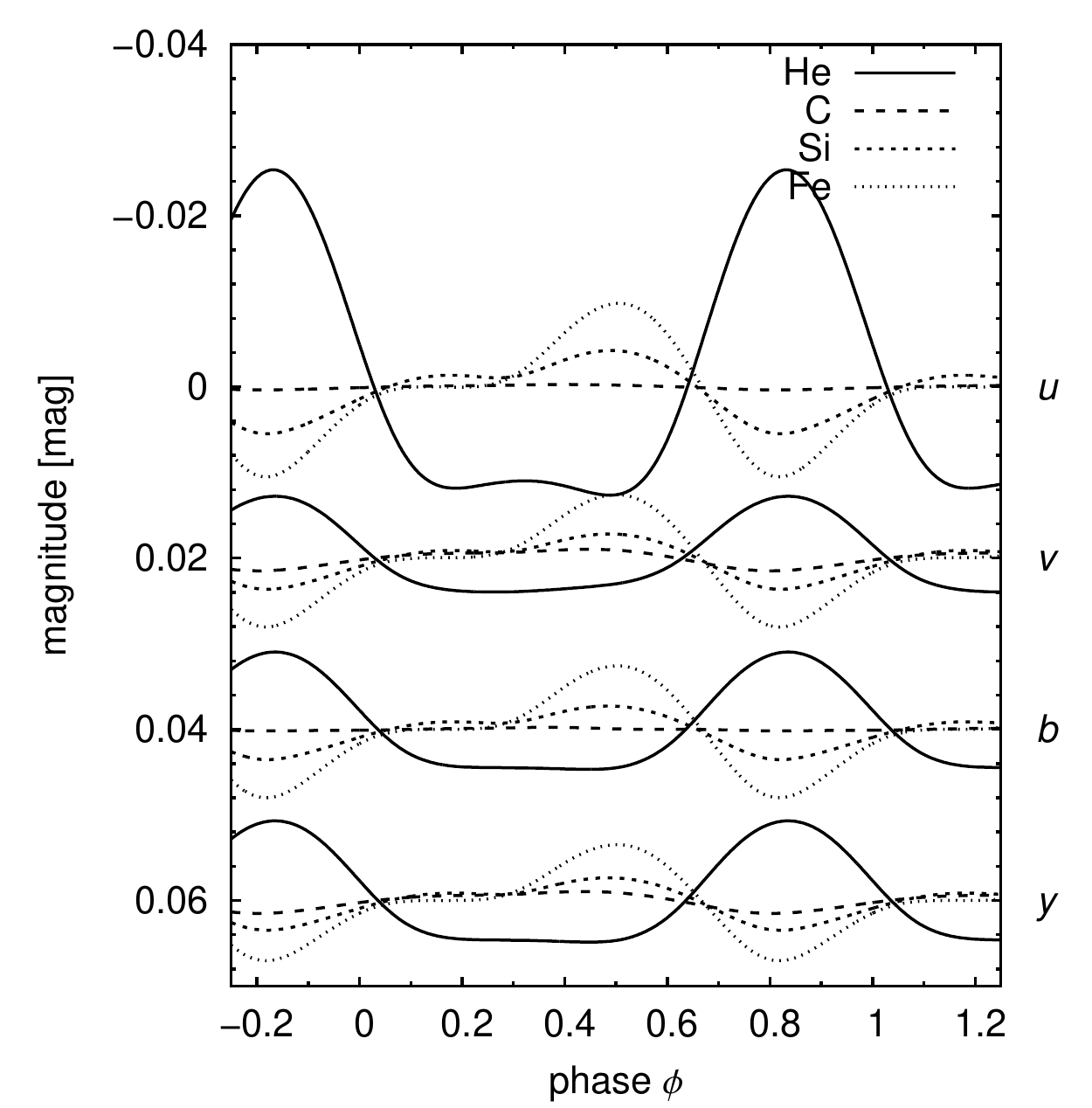}}
\caption{Predicted light variations of \hvezda\ in the Str\"omgren photometric
system calculated considering only the variations of individual elements. The abundance of
other elements was fixed. Light curves in individual filters were
vertically shifted to
better demonstrate the light variability.}
\label{prv_hvvel}
\end{figure}

The bound-free and bound-bound transitions of individual elements
can modify the temperature of model atmospheres.  This can
be seen in Fig.~\ref{tep}, where we compare the temperature distribution
of model atmospheres for typical overabundances of
\hvezda\,to the distribution for abundances slightly higher than solar.
The bound-free (due to ionization of helium, carbon, and silicon) and
bound-bound (line transition of iron) transitions absorb
the stellar radiation, consequently the temperature in the continuum
forming region ($\tau_\text{ross}\approx0.1-1$) increases with
increasing abundance of these elements. In the uppermost layers
$\tau_\text{Ross}\lesssim10^{-3}$, the stronger line cooling in the models with
enhanced abundance leads to the decrease of the atmosphere temperature. 

\begin{figure}
\centering
\resizebox{\hsize}{!}{\includegraphics{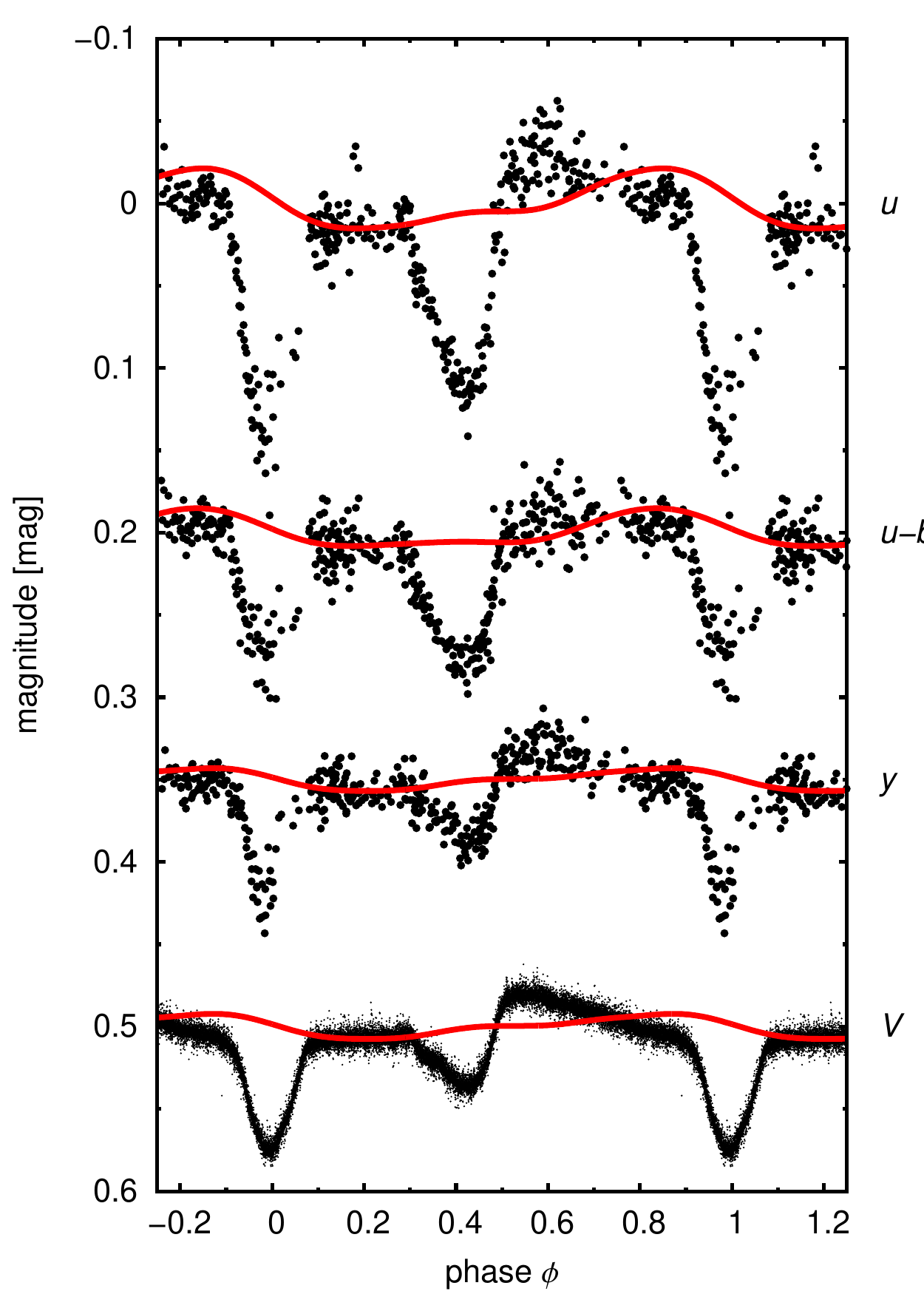}}
\caption{Predicted light variations of \hvezda\ (solid lines) computed taking
into account helium, silicon, carbon, and iron surface abundance distributions. 
Light curves in individual filters were vertically
shifted to better demonstrate the light variability.  The top three light curves are from the 
Str\"omgren photometric system, with the data again that of \citet{H77}.  
The bottom light curve is calculated for the wide-band visual filter corresponding to the MOST data \citep{town13}. }
\label{sigorie_hvvel}
\end{figure}

In atmospheres with overabundant helium, carbon, silicon or iron, enhanced
opacity leads to the redistribution of the flux from the short-wavelength part
of the UV spectrum to longer wavelengths, and also to the
visible spectral regions \citep[see Figure~\ref{prvtoky} and][]{Krt07}. Consequently,
overabundant spots are typically bright in the visual and near UV bands, and are
dark in the far UV bands. 

These flux changes can be detected in the variability of the SED and 
particularly in the variability of visible light. To demonstrate the optical variations,
we plot (Fig.~\ref{magtoky}) the relative magnitude difference
\begin{equation}
\label{tokmagroz}
\Delta m_\lambda=-2.5\log\zav{
\frac{H_\lambda(\varepsilon_\text{He},\varepsilon_\text{C},
\varepsilon_\text{Si},\varepsilon_\text{Fe})} {H_\lambda^\text{ref}}},
\end{equation}
against wavelength. Here $H_\lambda^\text{ref}$ is the reference flux calculated
for slightly overabundant chemical composition (with
$\varepsilon_\text{He}=-1.0$, $\varepsilon_\text{C}=-4.1$,
$\varepsilon_\text{Si}=-4.1$, and $\varepsilon_\text{Fe}=-4.7$). As can be seen
in Figure~\ref{magtoky}, the absolute value of the relative magnitude difference
is the largest in the $u$-band of the Str\"omgren photometric system, while it
is nearly the same in the visual bands, $v$, $b$, and $y$-band. The strong
maximum in the $u$-band model with enhanced helium is caused by the
filling of the Balmer jump. In the helium rich models, the jumps due to hydrogen
diminish, while the jumps due to helium become visible.

\subsection{Predicted visual light variations} \label{vislightcurve}

\begin{figure}
\centering \resizebox{\hsize}{!}{\includegraphics{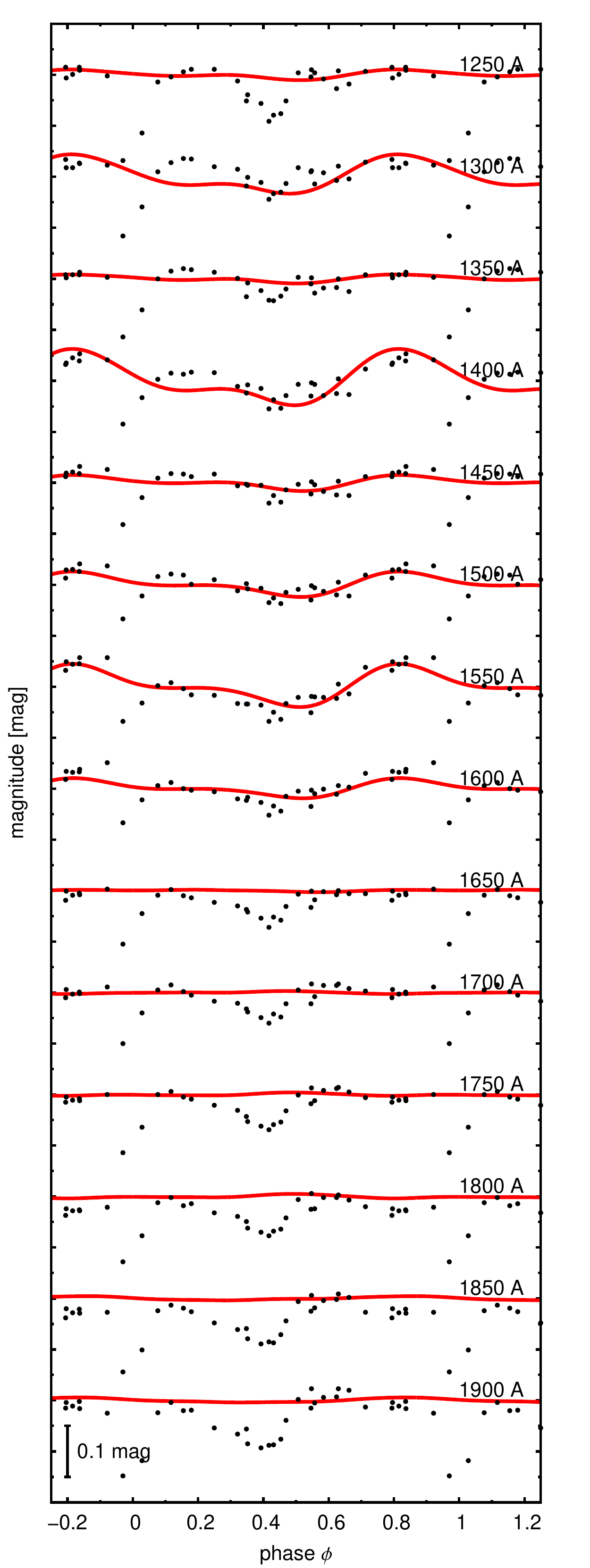}}
\caption{Comparison of the predicted (solid line) and observed (dots) UV light
variations for different wavelengths.
Curves for individual wavelengths were vertically shifted to
better demonstrate the variability.}
\label{hvveluv}
\end{figure}

Predicted light curves are calculated from the surface abundance maps
derived using MDI in Section 2 and from the emergent fluxes computed with the
SYNSPEC code, applying Eq.~\ref{velik} for individual rotational phases.

To study the influence of individual elements separately, we first calculated
the light variations with the abundance map of a single element
(Fig.~\ref{prv_hvvel}), assuming a fixed abundance of other elements
($\varepsilon_\text{He}=-1.0$, $\varepsilon_\text{C}=-4.1$,
$\varepsilon_\text{Si}=-4.1$, $\varepsilon_\text{Fe}=-4.7$). 
Fig.~\ref{prv_hvvel} shows that helium, iron, and silicon contribute predominantly to
the light variations, while the contribution of carbon is only marginal. This is
due to the large overabundance of these elements within the spots and 
their large abundance variations on the stellar surface. The amplitude of the light
variations is the largest in the Str\"omgren $u$-band, as expected from
Fig.~\ref{magtoky}. Because the overabundant regions are brighter in the $uvby$
colors, the predicted light variations reflect the equivalent width variations.
The light maximum occurs at the same phase at which the equivalent width of a
given element is the greatest. Helium lines are strongest for phase 0.8,
consequently the light curve due to helium shows its maximum at phase 0.8.
Iron and silicon lines are strongest for phase 0.6, consequently the light
curve due to both iron and silicon shows its maximum at phase 0.6.

Incorporating the surface distributions of helium, carbon, silicon, and
iron in the calculation of light curves (Fig.~\ref{prv_hvvel}), we obtain
the results given in Figure~\ref{sigorie_hvvel}. Helium, due to its large overabundance,
dominantly influences the light curve of \hvezda. This is the opposite behavior
seen for HD~37776, where silicon was the main cause of the light
variability \citep{Krt07}.  This is due to the higher silicon abundance in HD~37776 by
about 0.8~dex.

The observed visual light curve of \hvezda\ has a maximum at phase $\sim$0.6, when
the iron lines are the strongest (Fig.\ref{sigorie_hvvel}), however, the model
indicates the light curve has a maximum for the phase of about 0.8 when the
helium lines are the strongest. Consequently, the observations strongly disagree
with theory in the optical region, and the excess brightness at phase 0.6 cannot be explained 
as an effect of inhomogeneous surface abundance of considered elements.

\begin{figure*}
\centering \resizebox{\hsize}{!}{\includegraphics{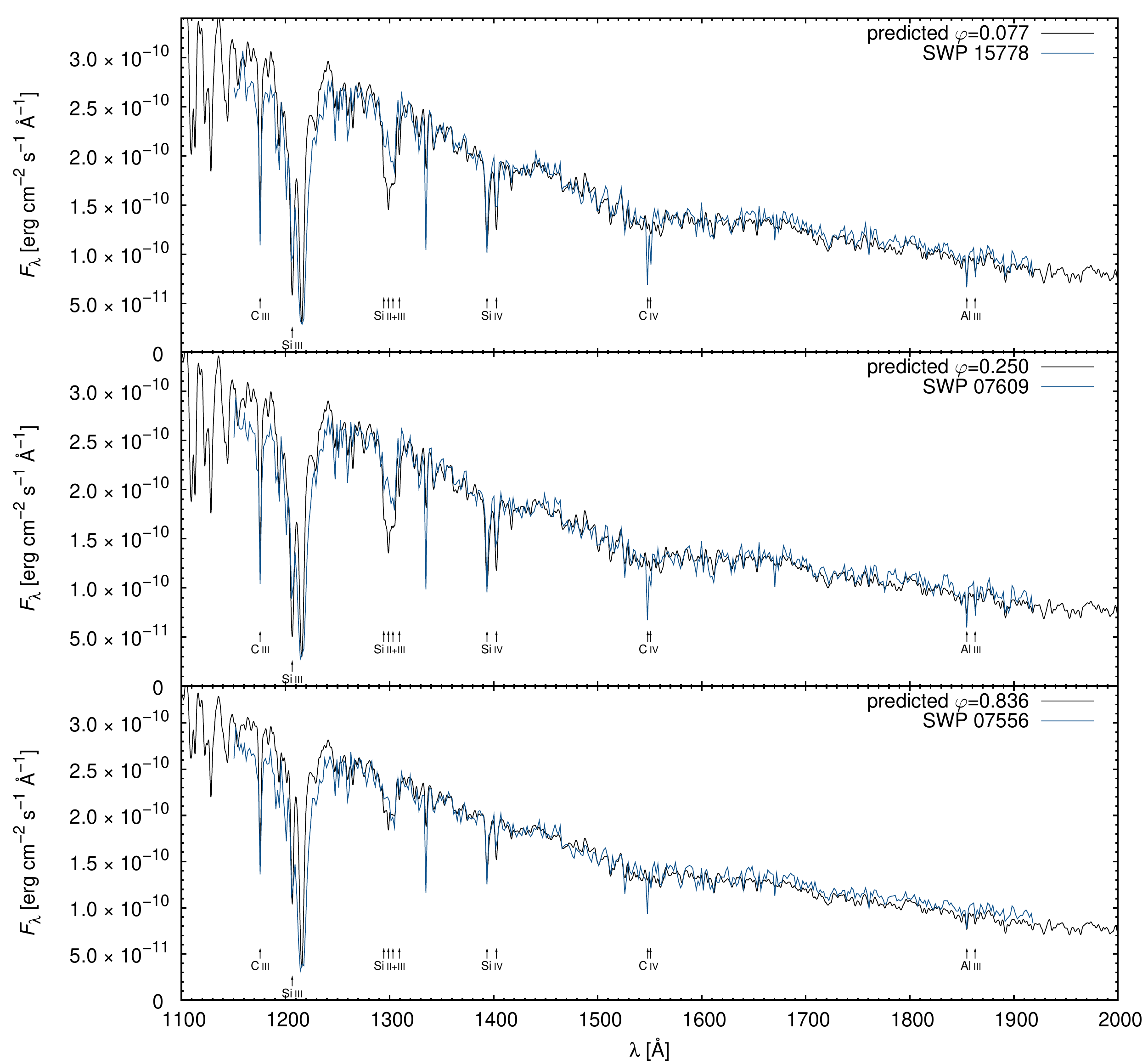}}
\caption{Comparison of observed (IUE) and predicted fluxes in the UV region for selected phases.  
The predicted flux was smoothed by a Gaussian filter with dispersion 1.3 \AA, which roughly corresponds the the IUE data.}
\label{sigoriefm}
\end{figure*}

\section{Extended analysis of the UV variability}
\label{uv}

The light variability caused by the uneven distribution of elements on the surface of \hvezda\ 
is connected with the flux redistribution from the far UV to the near UV and visible
regions (see e.g., Figure\,\ref{prvtoky}).  Consequently, the light variability in the far 
UV region provides an important test of the light variability mechanism.  
To test these predictions quantitatively, we extracted IUE observations of
\hvezda\ from the INES database \citep{ines} using the
SPLAT package \citep[see also \citealt{pitr}]{splat}. Here we use low-dispersion
large aperture spectra in the domain 1250--1900~\AA\ (SWP camera).

\subsection{Narrow band UV variations}

As a first comparison of the UV fluxes, we focused on narrow band
variations. The observed and predicted fluxes were both smoothed
with a Gaussian filter with a dispersion of $25\,$\AA. The resulting predicted
and observed UV variations are given in Fig.~\ref{hvveluv}.
Unlike the visible region, light curves calculated from the MDI-derived abundance
maps are able to explain at least the main trends of UV light variability of
\hvezda. At 1250\,\AA\ and 1900\,\AA\, light variability is dominated by
light absorption in the circumstellar environment. The observed light curve
is nearly constant in the periods between the occultations, with the 
calculated light curve also producing little variability. Conversely, there
are significant variations of the observed flux at 1400\,\AA\ and 1550\,\AA.
These variations can be relatively nicely reproduced by our models. The light
variability in these regions is mainly due to silicon and iron, while helium does not
significantly influence the light variability in the UV region.

\subsection{UV flux variations}

The variability of the UV spectral energy distribution in Figure~\ref{sigoriefm} is
only marginal. The predicted and observed emergent flux distribution agree
relatively well during ($\varphi=0.077$) and outside 
($\varphi=0.250$, $\varphi=0.836$) eclipses. The flux is greatly influenced by numerous
iron lines, whose accumulation around $1550\,$\AA\ causes flux depression in
this region. A detailed inspection reveals that the predicted flux is somewhat
higher than observed in the region $1150-1250\,$\AA\ for all phases.
Moreover, the predicted flux is slightly lower than observed in the
region $1650-1900\,$\AA\, during the maximum of the UV flux variability
($\varphi=0.836$). This may possibly be the source of disagreement between the
light curves in the optical region.

\subsection{UV line variations}

\begin{figure}
\centering \resizebox{\hsize}{!}{\includegraphics{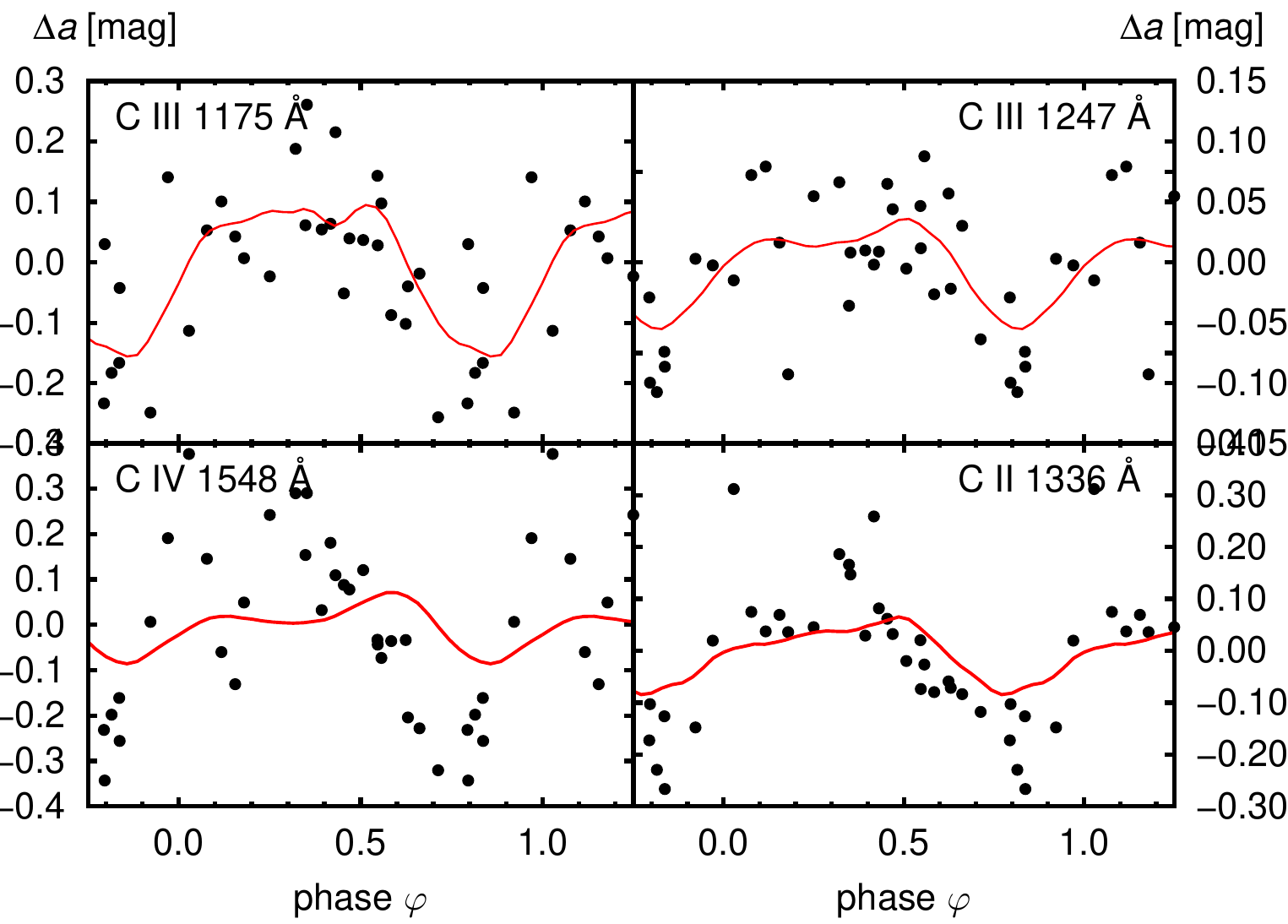}}
\centering \resizebox{\hsize}{!}{\includegraphics{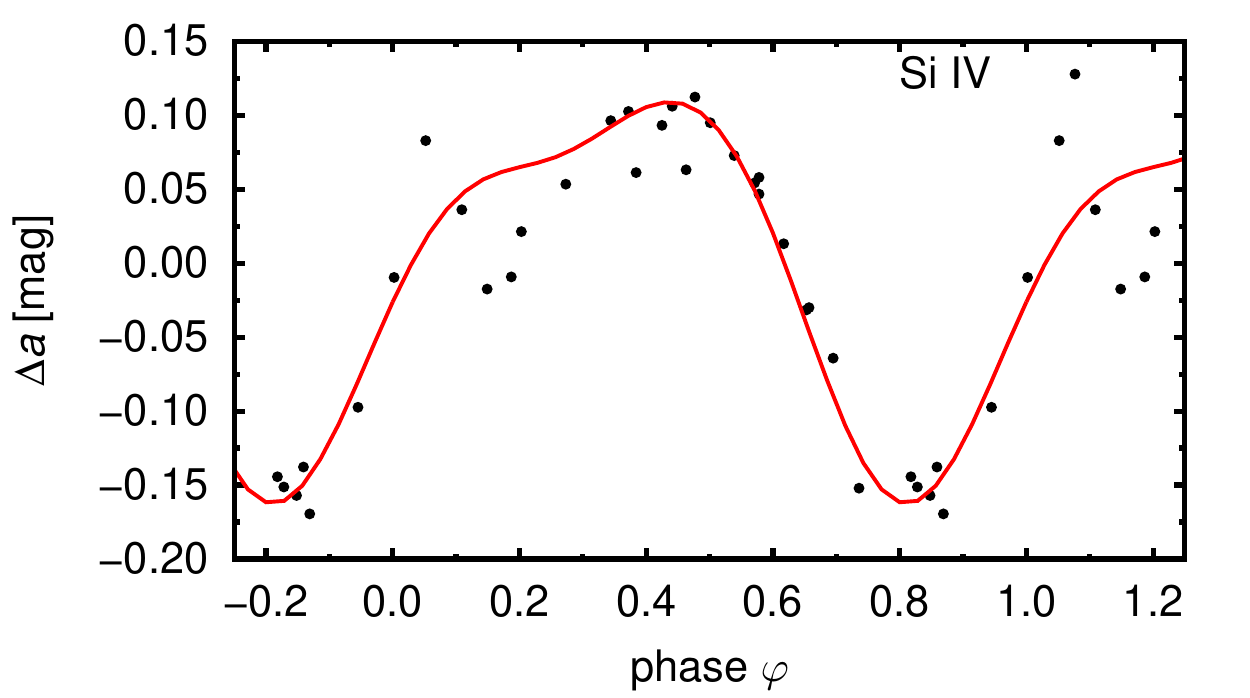}}
\caption{Observed variability of the line indices (dots) in comparison with the predicted indices (lines). Upper panel: Carbon line indices. Lower plot: Silicon line index. The line indices were calculated according to Eq. (10).}
\label{car}
\end{figure}

The shapes of many weak lines and the resonance lines of Si\,{\sc iv} at 
$\sim1400\,$\AA\ seem to be well reproduced by the models in
Figure~\ref{sigoriefm}. On the other hand, the strengths of strong lines of
Si\,{\sc ii} and Si\,{\sc iii} at 1270\,\AA\ and $1300\,$\AA\ and the
C\,{\sc iv} resonance lines are not reproduced by the models. We use the following indices 
to study line variability:
\begin{subequations}
\label{carin}
\begin{align}
a(\text{C\,{\sc ii} }
1336\,\text\AA)&=m_{1335.2}-\frac{1}{2}\zav{m_{1332.0}+m_{1338.75}},\\
a(\text{C\,{\sc iii} }
1175\,\text\AA)&=m_{1175.7}-\frac{1}{2}\zav{m_{1169.0}+m_{1181.25}},\\
a(\text{C\,{\sc iii} }
1247\,\text\AA)&=m_{1247.5}-\frac{1}{2}\zav{m_{1244.45}+m_{1249.15}},\\
a(\text{C\,{\sc iv} }
1548\,\text\AA)&=m_{1547.9}-\frac{1}{2}\zav{m_{1529.45}+m_{1557.95}},\\
a(\text{Si\,{\sc iv}})&=m_{1393.65}-\frac{1}{2}\zav{m_{1388.65}+m_{1398.7}}.
\end{align}
\end{subequations}
These indices were used in the analysis of \citet{Krt13} for HD\,64740, 
who modified a similar set of indices developed by \citet{Shore87}.
Figure~\ref{car} compares the observed and predicted variations of the line
indices with respect to their mean values. We are able to explain the variations
of many individual carbon and silicon lines. The disagreement between predicted
and observed line profiles of Si\,{\sc ii} and Si\,{\sc iii} could be caused
by an inhomogeneous vertical distribution of silicon in the atmosphere. The
observed line strengths of C\,{\sc iv} at $1548\,$\AA\ and $1551\,$\AA\ are
significantly larger than predicted. The observed asymmetry of these
lines with extended blue wings at each phase indicates formation in the wind.
The line strength variations result from the variations of wind mass-flux on
the stellar surface. Enhanced absorption in the C\,{\sc ii} line at 1336\,\AA,
C\,{\sc iv} lines, and Si\,{\sc iv} lines observed around the phase $\varphi\approx0$
possibly originates in the corotating magnetospheric clouds.

\section{Confrontation: Round 2}

In paper I, we first confronted the RRM model with new high-resolution spectropolarimetry of $\sigma$ Ori E to demonstrate that the field configuration assumed by \citet{town05} was incompatible with current observations.   This first step revealed that more comprehensive modeling was required to explain the intricate physical phenomena of this star.   While the initial success of that model is now attributed to a bug in the code, its inspiration has driven studies to better understand the intimate details of the interaction of rotation, magnetic field, and mass-loss, propelling the new wave of high quality data acquisition presented in this and other papers.  The current version of the RRM model now does a marginally good job of replicating the qualitative variations of the system, but several discrepancies appear to test the accuracy of the model, and the excess brightness in the optical photometry at phase 0.6 persists.     

The magnetic field topology in this application of the ARRM model is taken from the MDI analysis in Section~\ref{Bfield}, 
and is therefore consistent with the observations of longitudinal magnetic field.  
Several checks were performed to to ensure that the model adopted the same phasing and rotation 
as the derived maps, particularly after the discovery of the bug in the initial model.  
The H$\alpha$ emission comparison is shown in Figure ~\ref{fig:arrm-halpha}, and while the model 
reproduces much of the variability, there is one glaring inconsistency 
in the strength of the red-shifted cloud at phase $\sim$0.75.  If we carefully inspect the other features throughout the rotation, it seems this one aspect is the only serious problem, 
indicating the model is predicting too much emitting material in that ``cloud'' at that particular phase.   
Likewise, Figure ~\ref{fig:arrm-light} shows a similar reversal in strength of the the model photometric minima with the deeper minimum occurring at the opposite phase as compared to the observed light curve.   

Section~\ref{vislightcurve} demonstrates clearly there is no photometric resolution to the issue of the excess brightening in the second half of the rotation phase, or the irregular shape of the minimum at phase 0.4, both clearly seen in the MOST photometric light curve.
As modeling the photometric contribution to the brightness variability does not resolve the discrepancies, we must at this point consider that the ARRM model is not currently capable of explaining all of the observed variabilities of $\sigma$ Ori E and similar stars in detail, and that additional physics is necessary.

\subsection{Where is the missing physics?}

Where does this ``extra'' brightness come from?  If we look at Figure~\ref{fig:arrm-vis}, geometry could be a possible explanation.   Near phase 0.6-0.8, where we see brightening, the magnetosphere is seen more or less pole-on, at which we are able to view all of the emitting surface of the magnetospheric material.  At the phase 0.2, we see that the material is viewed more edge-on, with a much less direct view of most of the emitting surface.  It is possible (and likely) that the star's radiation may be reflecting off of the inner edge of the magnetosphere, thus creating a brightening effect.  At this iteration, this and other types of scattering are not considered in the RRM model, but likely contribute to the observed brightness.  This is extra physics that would need to be implemented into the next version of the code to diagnose the importance and type of scattering that could be responsible, i.e., Rayleigh scattering vs. electron scattering.

Another consideration is the shape of the ``clouds".  The ARRM model predicts, for $\sigma$ Ori E, two higher density concentrations within a larger disk-like structure.  However, the broadband polarization study of \citet{Carciofi} suggests that the material is configured in a much more concentrated structure, much more similar to dumbbells that a disk.  A more physically realistic density distribution may present different model predictions.   Another effect of the interaction of the star and the magnetic field is the way in which material leaves the surface of the star.  In a ``normal'' OB star, the mass-loss is purely dependent on the strength of the wind, and is proportional to the luminosity.  
Although the ARRM model considers effects from the magnetic field, the model still assumes standard CAK relations.  
The influence of strong field lines may modify the mass-loss process, changing the rate at which the material leaves the star resulting in higher or lower mass-loss.  
The complexity of the field structure may further constrict movement of mass from the surface.    
An effect like this could be responsible for the discrepancies between the observed H${\alpha}$ emission and the model prediction. 
Similarly, the influence of the surface abundance inhomogeneities may play an additional role in altering the mass-loss rate.

\section{Summary and future work}

In the first paper of this series, we evaluated the appropriateness of the magnetic field assumed by \citet{town05} in which they applied the RRM model to the physical case of $\sigma$ Ori E.
The conclusion of that work, that a more complex field was required, served as the motivation for the analyses contained in this subsequent paper.  We have used the same high-resolution 
spectra from Paper I to produce MDI maps of both the magnetic topology and the surface abundance distributions of He, C, Si, and Fe.  The derived magnetic topology, which can be represented by a superposition of a dipole plus quadrupole, is consistent with 
our previous assertion that the field contains higher-order components.  This magnetic configuration was then fed into the newly developed arbitrary magnetic 
field version of the RRM model to re-evaluate the consistency between the model predictions and the observed variations.  The resultant predictions qualitatively resemble the observed spectral and photometric variations, but fail to reproduce still several details.  While the magnetic field is fixed, the model now predicts a larger density in one of the clouds, which is observed, but located on the wrong side of the star.   The new model also does not resolve the previously-identified issues, specifically the excess brightness observed at phase $\sim$0.6.  

While clearly some of these issues are related to failings of the model, we proceeded to use the second result of the MDI modeling, the abundance maps, 
to determine the effect of the surface inhomogeneities on the photometric light curve.  With these maps as input, a synthetic light curve was computed with 
the aid of a grid of stellar atmosphere models.  The total resultant light curve, a product of the contributions from  He, C, Si, and Fe, provides no solution for 
the discrepancies between the model and the observations.  The light curve is dominated by the He contribution, as this element exhibits the largest variation of abundance 
across the surface, however, its maximum abundance is viewed at phase 0.8, too late to be helpful in accounting for the feature at phase 0.6.  An analysis of the 
UV light variability shows good agreement between observed variability and computed light curves, however, some narrow wavelength bands are affected by the 
eclipsing of the magnetospheric material, causing enhanced absorption at similar phases as in the optical. 

We thus conclude that the discrepancy between the models and the observations lies in the treatment of the stellar magnetosphere, and that the RRM 
model must undergo further revision.  Adding the ability to accommodate an arbitrary magnetic field topology is just the first step towards our greater understanding
 of this system, but there are multiple issues to be addressed in future iterations.  The inclination angle first and foremost should be determined using all the available 
 information.  The model will need to incorporate and test the effects of scattering processes so as to determine their ability to alter the synthetic observations.  
 Tests should be run to understand how the field interacting with the wind could change the stellar mass-loss.
When the RRM model was developed, massive star magnetism was just beginning to progress, and at the time, $\sigma$ Ori E was a unique object, the only known member of its class.  
However, in the past decade, research has grown rapidly, and there are a growing number of stars discovered to have similar properties, and exhibit similar variabilities.  
By optimizing the RRM model for $\sigma$ Ori E, we can create a blueprint to fully comprehend the physical properties, the interactions within and outside of the star, 
and the evolutionary implications for this fascinating group of stars.

\section*{Acknowledgments}

We thank the referee, Dr. J. Landstreet, for valuable comments, which improved the analysis and manuscript.  MEO acknowledges financial support from the NASA Delaware Space Grant, NASA Grant \#NNG05GO92H, and the postdoctoral program of the Czech Academy of Sciences.   OK is a Royal Swedish Academy of Sciences Research Fellow supported by grants from the Knut and Alice Wallenberg Foundation and the Swedish Research Council.  JK, MP, and ZM were supported by the grant GA \v{C}R P209/12/0217.  RHDT acknowledges support from NSF grants AST-0904607 and AST-0908688.  GAW acknowledges Discovery Grant support from the Natural Science and Engineering Research Council of Canada (NSERC), and from the Academic Research Program of the Royal Military College of Canada.    SPO acknowledges partial support from NASA ATP Grant \#NNX11AC40G. The computations presented in this paper were performed on resources provided by SNIC through Uppsala Multidisciplinary Center for Advanced Computational Science (UPPMAX).

\bsp

\label{lastpage}

\end{document}